\documentclass{aa}

\usepackage{graphicx}
\usepackage{txfonts}
\usepackage{siunitx}
\usepackage{xcolor}
\newcommand{\kms}{{km\,s}$^{-1}$}

\newcommand{\teff}{$T_\mathrm{eff}$\,}
\newcommand{\logg}{$\log g$\,}
\newcommand{\vt}{$v_{\rm{t}}$}
\newcommand{\vsini}{$v_{\rm{e}} \sin i$}

\begin{document}

   \title{The VLT-FLAMES survey of massive stars}
   \subtitle{NGC\,2004\#115: A triple system hosting a possible short period B+BH binary}
   \titlerunning{A possible B+BH binary in the triple system NGC\,2004\#115}

%   \subtitle{I. this is the subtitle}

   \author{D.~J. Lennon\inst{1,2},
        P.~L. Dufton\inst{3},
        J.~I. Villase\~nor\inst{4,5},
        C.~J. Evans\inst{6},
        N. Langer\inst{7,8},
        R. Saxton\inst{9},
        I.~M. Monageng\inst{10,11},
        S. Toonen\inst{12}
          }
\authorrunning{D.~J.~Lennon et al.} 

   \institute{Instituto de Astrof\'isica de Canarias,E-38\,200 La Laguna, Tenerife, Spain
        \and
        Dpto. Astrof\'isica, Universidad de La Laguna, E-38\,205 La Laguna, Tenerife, Spain
        \and
        Astrophysics Research Centre, School of Mathematics \& Physics,  Queen’s University, Belfast, BT7 1NN, UK 
        \and
         Institute for Astronomy, University of Edinburgh, Blackford Hill, Edinburgh EH9 3HJ, UK
        \and
        Institute of Astronomy, KU Leuven, Celestijnenlaan 200D, 3001, Leuven, Belgium 
        \and
          UK Astronomy Technology Centre, Royal Observatory, Blackford Hill, Edinburgh, EH9 3HJ, UK
         \and
         Argelander-Institut f\"ur Astronomie, Universit\"at Bonn, Auf dem H\"ugel 71, 53121 Bonn, Germany
         \and
         Max-Planck-Institut f\"ur Radioastronomie, Auf dem H\"ugel 69, 53121 Bonn, Germany
         \and
         Telespazio UK for ESA, European Space Astronomy Centre, Operations Department, 28691 Villanueva de la Cañada, Spain
         \and
         South African Astronomical Observatory, P.O Box 9, Observatory, 7935 Cape Town, South Africa
         \and
         Department of Astronomy, University of Cape Town, Private Bag X3, Rondebosch 7701, South Africa
         \and
         Anton Pannekoek Institute for Astronomy, University of Amsterdam, 1090 GE Amsterdam, The Netherlands
             }

   \date{}

% \abstract{}{}{}{}{} 
% 5 {} token are mandatory
 
  \abstract
  %{} leave it empty if necessary 
   {NGC\,2004\#115 was classified as a single lined (SB1) Be spectroscopic binary in the Large Magellanic Cloud. Its H$\alpha$ morphology is reminiscent of the Galactic systems LB-1 and HR\,6819, both of which are proposed as either Be+BH or Be+stripped He-star systems.}
  {Multi-epoch optical spectra of NGC\,2004\#115 are used to determine if this binary can be explained by either of these two scenarios, and hence shed additional light on these interesting systems.} 
   {VLT-FLAMES and SALT-HRS data covering a baseline of $\sim$20 years were analyzed to determine radial velocities and orbital parameters, while non-LTE model atmospheres were used to determine stellar parameters and the relative brightness of the system components. Archive MACHO, $Gaia$, and XMM-Newton data provide additional constraints on the system.}
   {NGC\,2004\#115 is found to be a triple system consisting of an inner binary with a period $P$=2.92\,d, eccentricity $e\sim$0.0, and mass function $f$=0.07\,$M_\odot$. The only firmly detected star in this inner binary is a B2 star, the primary, with a projected rotational velocity (\vsini) of 10\,\kms\ and a luminosity of $\log L/L_\odot$=3.87. It contributes $\sim60\%$ of the total $V$-band light, with the tertiary contributing the other $\sim40\%$ of the light, while the secondary is not detected in the optical spectrum. The possibility that the primary is a low mass inflated stripped star is excluded since its Roche radius would be smaller than its stellar radius in such a compact system.  A main sequence star of mass 8.6\,$M_\odot$ is therefore inferred; however, the assumption of synchronous rotation leads to a secondary mass in excess of 25\,$M_\odot$, which would therefore be a black hole.
   The tertiary is detected as a fainter blended component to the hydrogen and helium lines, which is consistent with a slightly less massive B-type star, though with \vsini$\sim$300\,\kms. The data do not permit the characterization of the outer period, though it likely exceeds 120 days and is therefore in a stable configuration. 
   The disk-like emission is variable, but may be associated with the inner binary rather than the rapidly rotating tertiary.
   XMM-Newton provides an upper limit of $5\times10^{33}$\,ergs\,s$^{-1}$ in the 0.2--12\,keV band which is consistent with, though not constraining, the system hosting a quiescent B+BH binary. A number of caveats to this scenario are discussed.
   }
   {}


   \maketitle
%
%-------------------------------------------------------------------

\section{Introduction}

The proposed stellar mass black hole (BH) candidates in the long period binary LB-1 \citep{liu} and the triple system HR\,6819 \citep{rivinius} are possible additions to the class of X-ray quiet Be+BH systems of which MWC\,656 \citep{casares} is the only confirmed member to date. These systems, with periods of some tens of days, are especially interesting as theory \citep{langer} predicts that they may contribute to the population of BH+BH systems that led to the discovery of gravitational waves \citep{abbott}. 
However, the nature of these new systems is disputed \citep{abdulmasih,elbadry2021,sergio2020,irrgang,shenar,bodensteiner2020,lennonlb1}.

Morphologically, the spectra of LB-1 and HR\,6819 are composite, having a narrow-lined B-type spectrum superimposed on a relatively featureless Be-type spectrum from which only the hydrogen Balmer and stronger \ion{He}{i} lines are detected.  The B-type spectrum displays clear radial velocity variations of some tens of \kms\ consistent with a long-period binary, while the Be-type spectrum is almost stationary to within a few \kms.  
Interpreting the latter as binary motion anticorrelated with that of the B-type star leads to mass ratios on the order of 10, and the systems' natures then hinge on the derived masses of one or another of the components.
If the B-type star is a main sequence star, then the Be-type spectrum may signify a candidate BH or neutron star (NS) plus accretion disk \citep{liu,abdulmasih}, though tension may arise if the mass of the B-type star is small enough such that the implied reflex motion of the BH or NS accretion disk becomes larger than observed \citep{sergio2020}.
Conversely, if the Be-star is a classical Be star, then the B-type star may be a stripped helium star contracting to become a hot subdwarf \citep{irrgang,shenar,elbadry2021,bodensteiner2020}.
Finally, the Be-like emission can be interpreted as being stationary \citep{elbadry2020,abdulmasih} and unrelated to the secondary of the B-type star, either belonging to a third very long period Be-star in the system \citep{rivinius}, or even as chance alignment with a Be-star \citep{safarzadeh}. In this case, the secondary to the B-type star is not identified in the spectrum. 
In the context of triple systems hosting BHs, HD\,96670 may also be relevant as \citet{gomez} suggest it may be a triple system with an inner short-period ($\sim$5.3\,d) binary consisting of a 23\,$M_\odot$ O-star plus a 6\,$M_\odot$ BH. 

The star considered here, with the Simbad designation Cl* NGC\,2004 ELS 115 (hereafter NGC\,2004\#115), lies on the periphery of the young cluster NGC\,2004, on the northern edge of the Large Magellanic Cloud (LMC). It was observed as part of the VLT-FLAMES Survey of Massive Stars \citep[FSMS,][]{evans2005,evans2006} and classified as a B2e star (without a luminosity classification). It was noted to be a single-lined spectroscopic binary (SB1) with double-peaked H$\alpha$ emission that also has a central absorption component. 
The star was subsequently discussed by \citet{dunstall2011} in their study of nitrogen abundances in Be stars from the FSMS, where they noted it as a short-period system and likely not a classical Be star. They gave its projected rotational velocity (\vsini) as 15\,km/s, which is close to the spectral resolution of the data, but it was not considered further.

\begin{figure}
\centering
\includegraphics[width=\hsize]{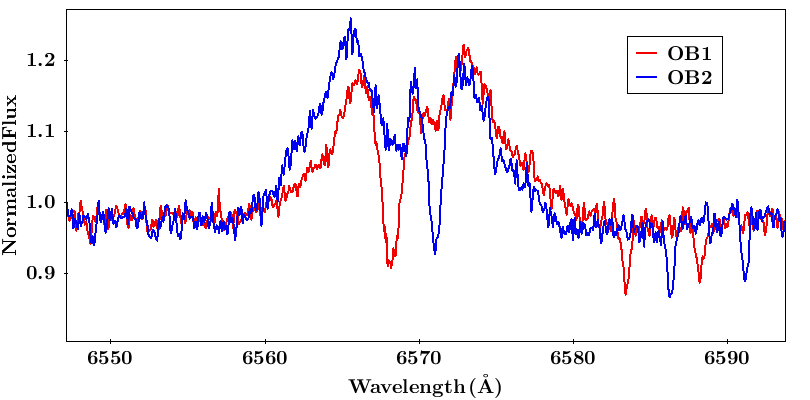}
\caption[]{Composite nature of the H$\alpha$ profile that consists of a broad double-peaked emission feature, plus a narrow absorption component from the B-type star, the narrow lines of the C\,{\sc ii} 6578/6582\,\AA\ doublet are further to the red. The two epochs from the VLT-FLAMES observations shown, OB1 (red) and OB2 (blue), illustrate significant velocity variability of the narrow components but a relatively stable broad H$\alpha$ emission. 
}
\label{fig:halpha}
\end{figure}

However, NGC\,2004\#115 recently attracted our attention as the morphology of its spectrum is strikingly similar to that of LB-1 and HR\,6819. It has a narrow-lined B-type spectrum that is clearly undergoing short-period (a few days) radial velocity variations. There is no clear evidence for a second spectrum other than the H$\alpha$ emission line that is almost stationary but it may exhibit small antiphase variations in the two epochs obtained (see Fig.\ref{fig:halpha} as well as subsection 3.3 and Fig.~\ref{fig:halpha_velocity} for a further discussion of H$\alpha$). However, additional observations were recently obtained with the Southern African Large Telescope (SALT; Buckley et al. 2006) to improve phase coverage and revealed the H$\alpha$ emission to be much weaker, and provide strong evidence that NGC\,2004\#115 is a triple system.

\section{Observational data}

The FSMS data are presented in detail by \citet{evans2006}, therefore we only summarize the relevant points here. NGC\,2004\#115 was observed with VLT-FLAMES between late 2003 and early 2004 with each of six high resolution grating settings; HR02 (central wavelength 3958\AA), HR03 (4124\AA), HR04 (4297\AA), HR05 (4471\AA), HR06 (4656\AA) and HR14 (6515\AA)\footnote{Now referred to as HR14A at the observatory.}, with wavelength coverage of each region listed in Table\,\ref{table:A.1}. All regions were observed in two observing blocks (OB1 and OB2), each of which was split into 3 contiguous 2275\,s exposures, for a total of 36 spectra with a typical resolution of $\sim$20\,000\footnote{The FSMS data analyzed here may be downloaded from https://star.pst.qub.ac.uk/\textasciitilde sjs/flames/}. 
However the signal-to-noise (s/n) ratio for the second OB of the HR03 setting was poor for all stars in this configuration, with only the first observation proving to be useful.

NGC\,2004\#115 was also observed with SALT using the High Resolution Spectrograph (HRS; Bramall et al. 2010, 2012; Crause et al. 2014). 
The four observations (17, 19, 22, and 23 September 2020) were performed in low resolution mode (R$\sim$14\,000) with exposure times of 2400\,s and covered a wavelength range of approximately 4000--8800 \AA. 
Primary reductions were executed with the SALT science pipeline (Crawford 2015) which include overscan correction, bias subtraction and gain correction. The subsequent reduction steps, which include background subtraction, arc line identification, removal of the blaze function, and merging of orders, were carried out with the \textsc{midas feros} (Stahl et al. 1999) and \textsc{echelle} (Ballester 1992) packages. 
A detailed description of the data reduction procedure is provided in Kniazev et al. (2016). Orders containing the broad H$\alpha$ and H$\beta$ lines were reduced following the recipe discussed in Appendix A of \citet{sergio2020}.

\section{System properties}

\begin{figure*}
    \centering
    \includegraphics[width=0.95\hsize]{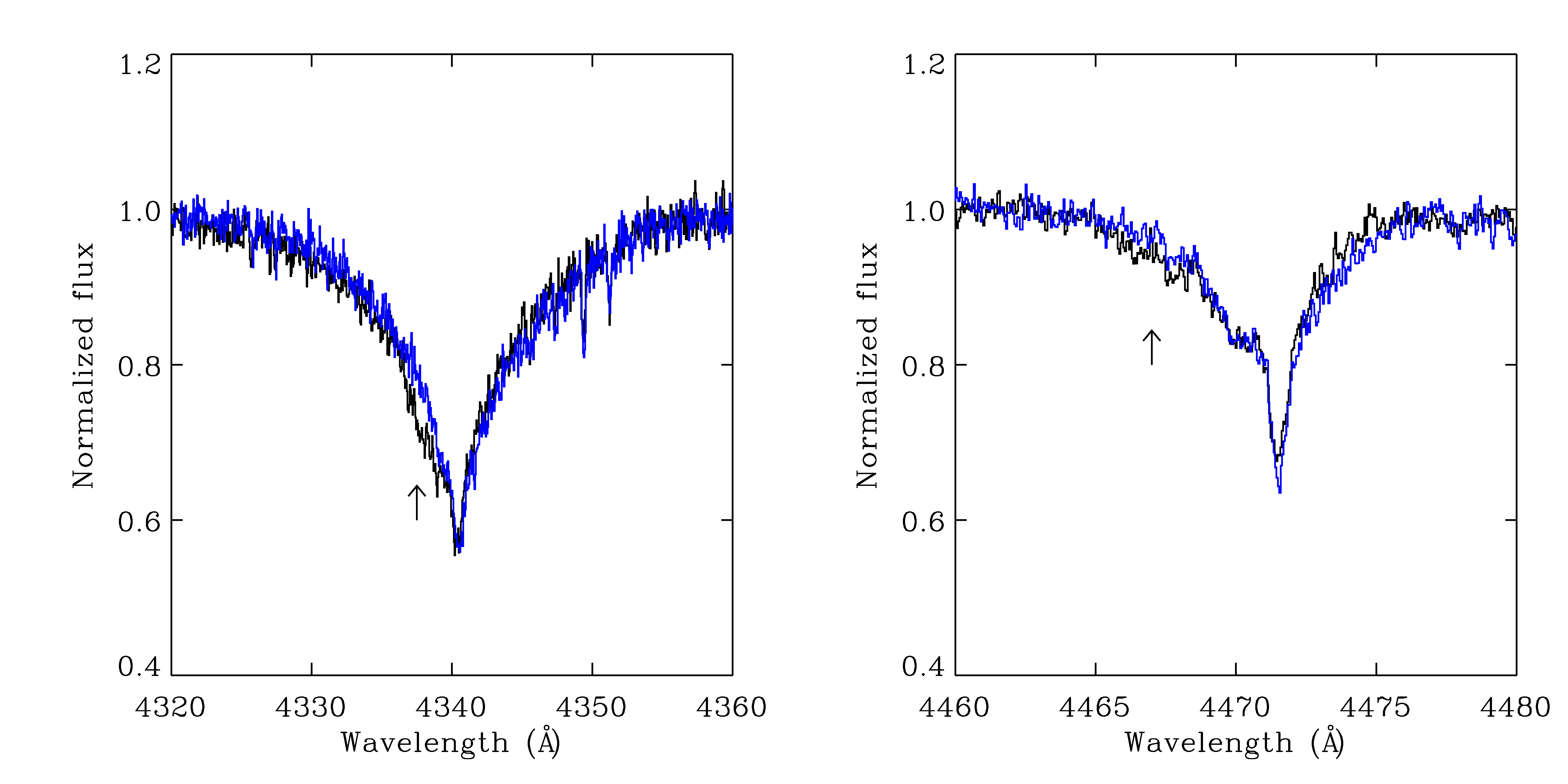}
    \caption{Left panel: Comparison of H$\gamma$ profiles from two VLT-FLAMES epochs in the rest frame of the narrow-lined B-type star. At these epochs the star is close to each quadrature phase.  An arrow indicates the extra absorption that we attribute to the presence of a third star, not present in the blue spectrum. If this component was due to a faint companion to the B-type star then it should be visible on the red wing of the blue spectrum. Instead its velocity at that epoch is almost identical to that of the B-type star, see subsection 3.2, hence no absorption is apparent at that epoch. Right panel: As for the left panel except here we plot two VLT-FLAMES epochs of the \ion{He}{I} 4471\,\AA\ line. Similar considerations apply.}
    \label{fig:tertiary}
\end{figure*}

The data for this system exhibit the spectral signatures of three components; a narrow lined B-type spectrum of spectral type B2 (illustrated in Fig.~\ref{spectra:primary}), an emission-line Be disk-like spectrum characterized by Balmer emission that is variable (no other emissions lines are detected at the present s/n), and the faint signature of another star in wings of the Balmer and \ion{He}{I} lines, as shown in Fig.~\ref{fig:tertiary}. We tentatively identify this last component as belonging to a third star, and suggest that the close companion to the narrow-lined star is undetected.
For convenience we refer to the narrow-lined B-type stars as the primary (sometimes as m$_1$), its undetected close companion as the secondary (m$_2$), and the faintly detected star as the tertiary (m$_3$). The source of the emission line spectrum we refer to as the disk. We refer to the primary+secondary system as the inner binary (at times the binary) and the inner binary plus tertiary as the outer binary.

%\begin{center}
%{\scriptsize
\begin{table}
\caption{Orbital parameters for the narrow-lined B-type star in NGC\,2004\#115; period ($P$), time of periastron ($T_p$), eccentricity ($e$), argument of the periastron ($\omega$) (as $e$ is small, $T_p$ and $\omega$ are not well constrained), systemic velocity ($\gamma$), and the amplitude or the radial velocity ($K_1$). Derived quantities are the semi-major axis distance to the center of mass ($a_1$) times sin$i$, and the mass function $f$. }
\centering
\begin{tabular}{lr}
\hline
\hline
Parameter			&Value\\
\hline
\multicolumn{2}{c}{Adjusted Quantities}\\
\hline
$P$ (d)		&2.918 $\pm$ 0.010\\
$T_p$ (HJD)		&2453005.45 $\pm$ 0.10\\
$e$			&0.02 $\pm$ 0.02\\
$\omega$ (deg)		&171 $\pm$ 10\\
$\gamma$ (km/s)	&326.5 $\pm$ 2.1\\
$K_1$ (km/s)		&62.4 $\pm$ 0.5\\
\hline
\multicolumn{2}{c}{Derived Quantities}\\
\hline
$a_1\sin i$ ($R_\odot$)	&3.60 $\pm$ 0.03\\
$f(m_1,m_2)$ ($M_\odot$)	&0.073 $\pm$ 0.005\\
\hline
\end{tabular}
\label{table:orbit}
\end{table}
%}
%\end{center}

   \begin{figure*}
   \centering
   \includegraphics[width=0.33\hsize]{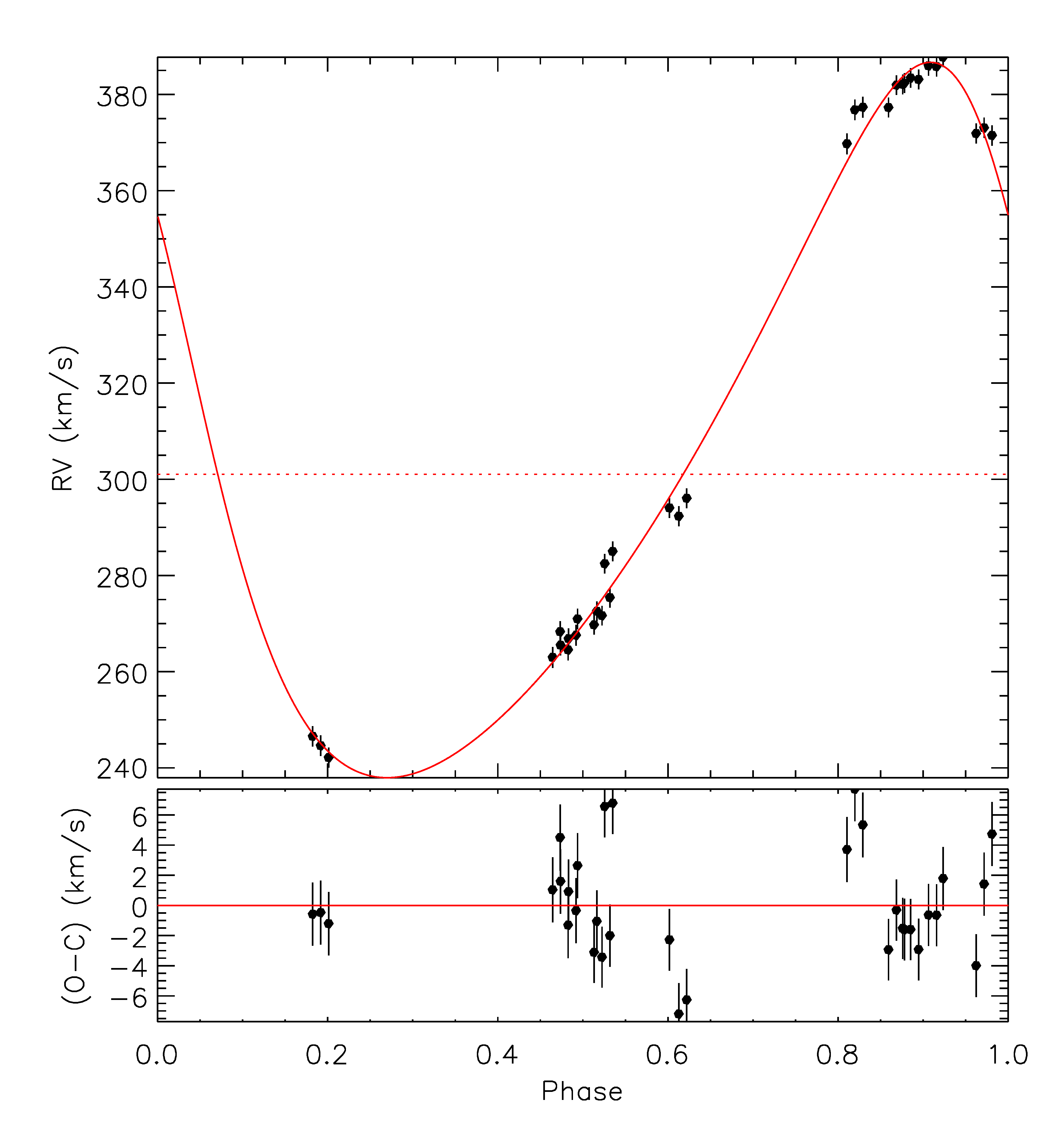}
   \includegraphics[width=0.33\hsize]{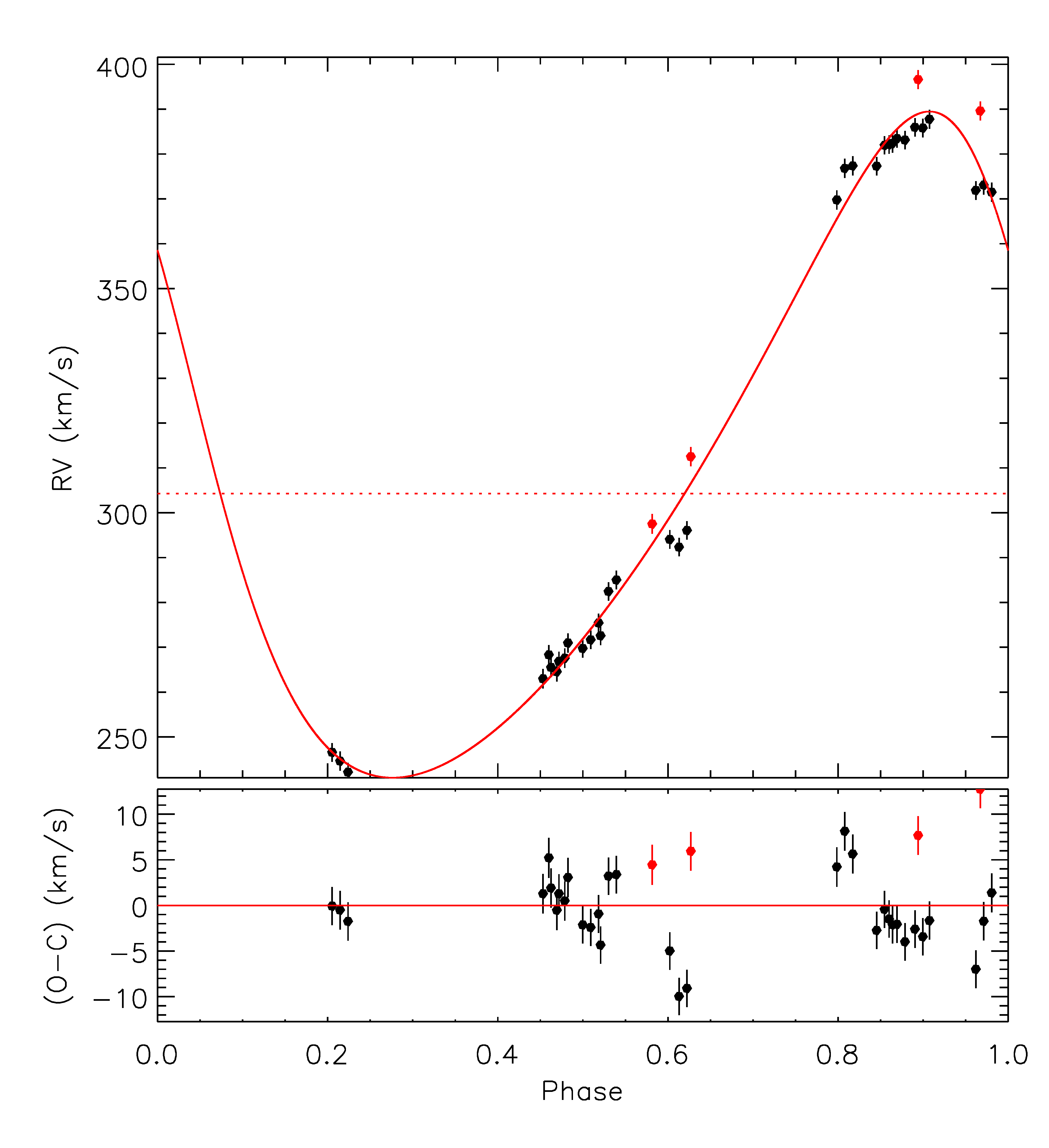}
   \includegraphics[width=0.33\hsize]{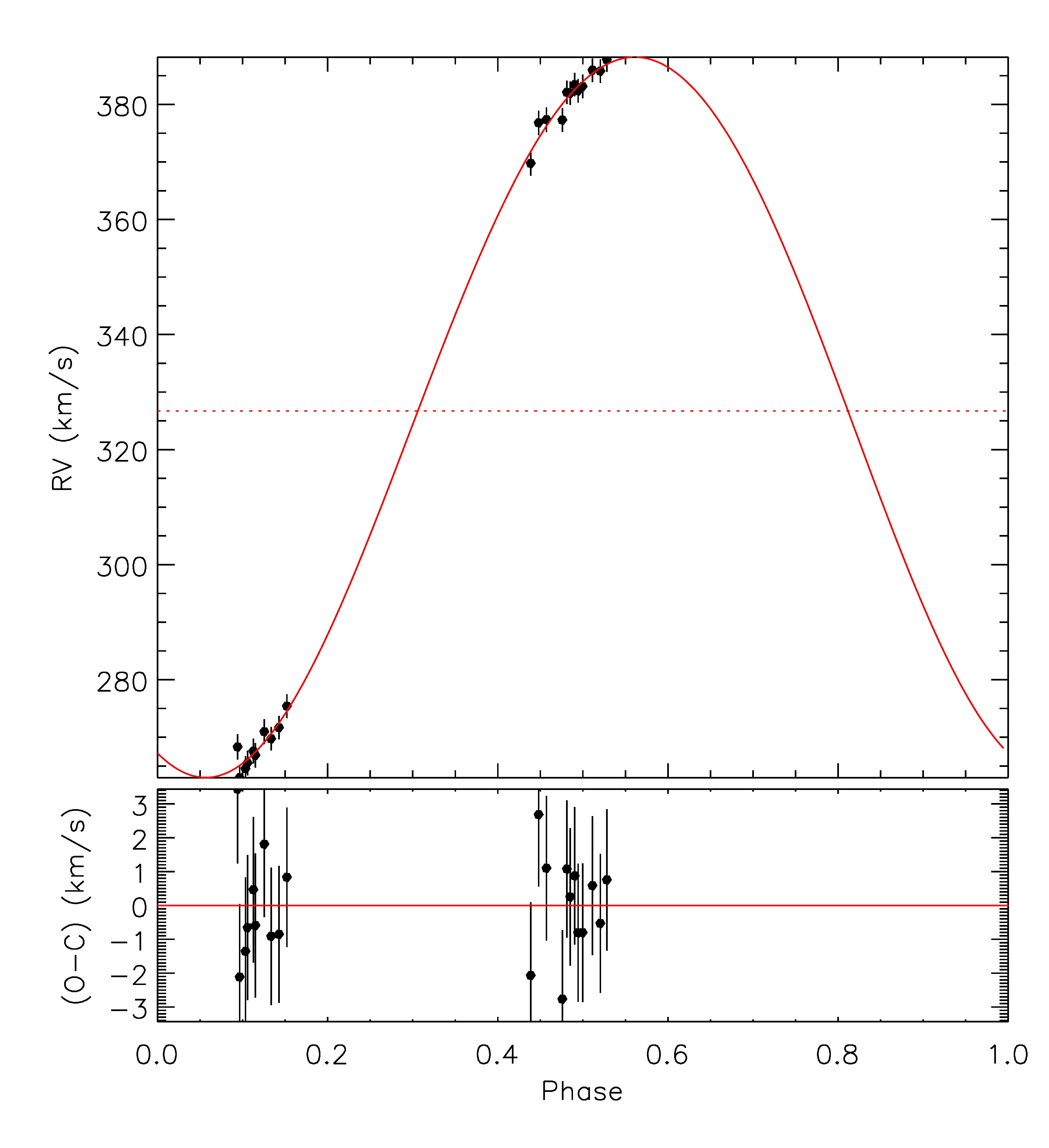}
      \caption{Radial velocity (RV) orbital solution for NGC\,2004\#115 based on velocity estimates from the metal lines in its spectrum. The left-most panel illustrates the fit using the VLT-FLAMES data only. The central panel shows how the fit, now determined using both VLT-FLAMES (black points) and SALT-HRS (red points) data, deteriorates when we include both datasets. The rightmost panel is the fit to the VLT-FLAMES data taken within the final week of that observing campaign, and fit parameters for this solution are listed in Table~1. The scales differ from panel to panel, in particular note the change in the scale of the O$-$C plots.
              }
         \label{fig:rvcurve}
   \end{figure*}

\subsection{Properties of the narrow-lined B-type star}

\subsubsection{Radial velocity analysis}

The sharpness of the metal lines in the primary enables precise measurement of its radial velocity, described in Appendix~\ref{appendix:rv}, with the results listed in Table \ref{table:A.2}.
The VLT-FLAMES velocities were initially used as inputs to the adaptive simulated annealing code {\tt rvfit} \citep{rvfit} to solve for the orbital parameters of the primary. This resulted in the determination of orbital parameters with period ($P$) of 2.88 days, eccentricity ($e$) of 0.27, and systemic ($\gamma$) and semi-amplitude ($K_1$) velocities of 301 and 74\,\kms\ respectively. The residuals were small, less than 6\,\kms, although the 1-$\sigma$ errors on the radial velocities were estimated as $\sim$1\,\kms. However addition of the SALT data resulted in a similar solution but with worse residuals, up to $\sim$10\,\kms, and with the overall rms of the fit increasing from 3.5 to 4.6\,\kms. This prompted a more careful examination of the VLT-FLAMES data. Noting that approximately two thirds of  these data were taken within a one week period a solution was sought with these data only. This resulted in a much better fit to the data, though with negligibly small eccentricity.
The {\sc rvfit} orbital solutions for these three cases are shown in  Fig.~\ref{fig:rvcurve}, while the detailed results for this last solution are listed in Table~\ref{table:orbit}.

\begin{figure*}
    \centering
    \includegraphics[width=1.0\hsize]{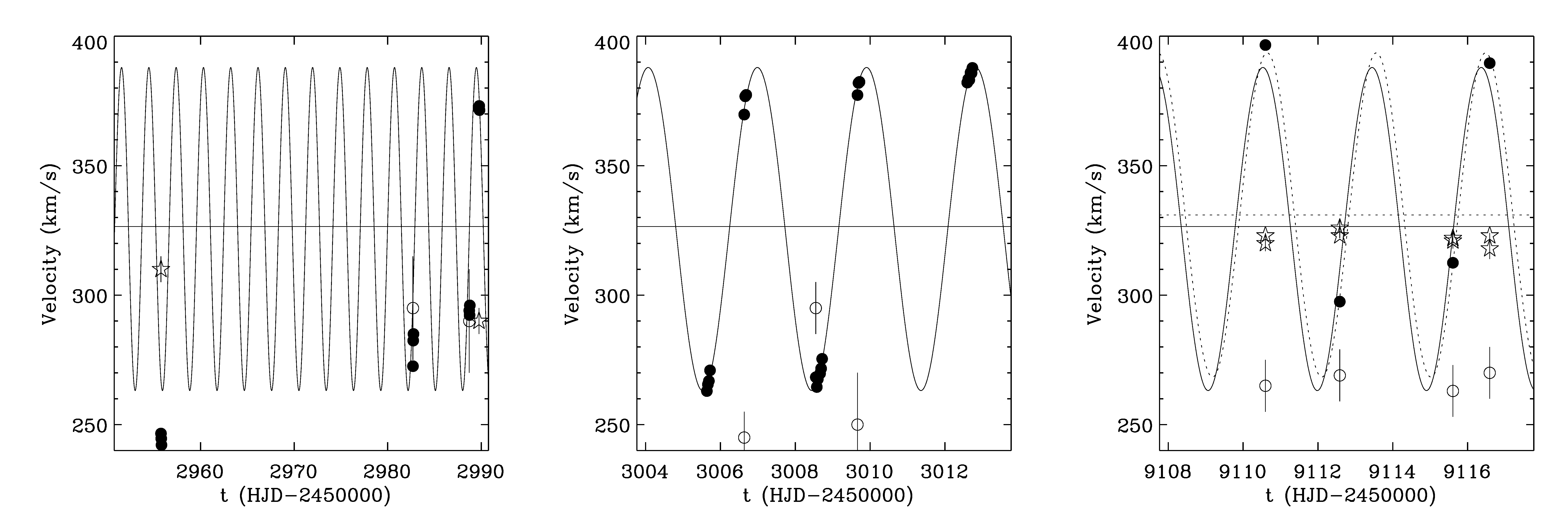}
    \caption{Central panel: Comparison of the solution derived from the late epoch FSMS data only (solid line) with the observed radial velocities of the primary, filled circles plus (small) error bars. Left panel: comparison of that solution with the early epoch FSMS data. Right panel; comparison with the SALT/HRS data. Horizontal lines indicate the systemic velocity of this solution. The dotted lines in the right panel are the solutions obtained by fitting those radial velocities but fixing both the period, eccentricity and semi-amplitude velocity to the values derived from the late FSMS data. Open star symbols are the velocities of the emission features (Table\,\ref{table:emrv}), and open circles are velocities of the broad-lined component (Table\,\ref{table:broadrv}) that we identify as a possible tertiary.}
    \label{fig:solutions}
\end{figure*}

Inspection of the fit to the early VLT-FLAMES and SALT-HRS data hints at a modulation of the radial velocity curve on longer timescales, though with small amplitude, as illustrated in Fig.~\ref{fig:solutions}. An obvious explanation is that NGC\,2004\#115 is a triple system, and indeed there is additional evidence from the SALT-HRS data of a blended component to the H$\beta$ line (discussed below). From the radial velocity discrepancies the semi-amplitude velocity for this tertiary component appears to be at least $\sim$20\,\kms, while the trend of the VLT-FLAMES data (early and late) over a 60 day time-span has no obvious minimum or maximum arguing for a period in excess of 120 days. However, given the sparseness of the sampling of the outer binary these are extremely tentative statements.

\subsubsection{Atmospheric analysis}

The grid of non-LTE (NLTE) models used in this analysis was calculated with the {\sc tlusty} and {\sc synspec} codes \citep{hubeny1988, hubeny1995, lan07} and adopted the ``classical'' model atmosphere assumptions, that is plane-parallel geometry, hydrostatic equilibrium, and that the optical spectrum is unaffected by winds. They also assumed a normal helium to hydrogen ratio (0.1 by number of atoms) and element abundances appropriate to the LMC. Microturbulence was not included as a pressure term in the hydro-static equilibrium equation. Further details can be found in \citet{rya03} and \citet{dufton2005, duf18},
while the analysis methods are discussed further in Appendix~\ref{appendix:atmosphere}.

\begin{table}
	\caption{Atmospheric parameters and abundances relative to hydrogen (in dex, n denoting the number of lines used per ion) for the B2 primary. Results are provided for the best fitting fractional contribution of 0.58 from the B-type star, and for comparison we also show results for a contribution of 1.0. The LMC baseline abundances are from \citet{hun07}.}             % title of Table
	\centering                          % used for centering table
	\begin{tabular}{l r r r c}        % centered columns (4 columns)
		\hline\hline                 % inserts double horizontal lines
		& \multicolumn{2}{c}{Fractional contribution} & & LMC \\ \cline{2-3}
		& 1.0  & 0.58 & n  & baseline \\    % table heading 
		\hline                                                   % inserts single horizontal line
		\teff (kK)             & 21.0$\pm$1.0       & 22.6$\pm$1.0  & -- & --   \\ 
		\logg (cm\,s$^{-2}$)   & 3.60$\pm$0.2       & 4.0:  & -- & --   \\
		\vt (\kms)             & 0$^{\dagger}$      & 0$^{\dagger}$ & -- & --   \\
		\vsini (\kms)          & 10$^{\dagger}$           & 10$^{\dagger}$      & -- & --   \\
		Radius ($R_\odot$)     & 7.8$\pm0.3$            & 5.6$\pm0.3$      & -- & --   \\
		BC (mag)               & $-2.09$            & $-2.27$       & -- & --   \\
		log$L/L_\odot$         & 4.03$\pm0.04$               & 3.87$\pm0.05$      & -- & --   \\
		C                      & 7.25$\pm$0.16      & 7.74$\pm$0.17 & 5  & 7.75 \\
		N                      & 6.83$\pm$0.12      & 7.12$\pm$0.10 & 4  & 6.90 \\
		O                      & 8.09$\pm$0.12      & 8.49$\pm$0.14 & 16 & 8.35 \\
		Mg                     & 6.63               & 7.22          & 1  & 7.05 \\
		Si                     & 6.71$\pm$0.12      & 7.25$\pm$0.07 & 3  & 7.20 \\
		%\\
		%R$_1$ ($R_\odot$)          & 7.7 & 6.9  &  &  \\
		%M$_1^{spec}$ ($M_\odot$)          & 8.5 & 13.9 &  &  \\
		%M$_2^{min}$ ($M_\odot$)    & 2.2 & 3.2 &  & \\
		\hline                                   %inserts single line
		\multicolumn{5}{l}{$\dagger$: \vt$\leq$2\,\kms\ and \vsini$\leq$15\,\kms}
	\end{tabular}
	\label{table:t_abund}
\end{table}
% Abundances checked by PLD 24/8/21

As already noted, there is little evidence of a second spectrum present other than the Balmer emission, and potentially broad blended Balmer lines, both of which are discussed below. However, the sharp metal lines are well observed except for the veiling effect of other sources. Indeed a straightforward analysis of the spectrum assuming no veiling leads to a solution in which all elements are underabundant (discussed below and illustrated in Table\,\ref{table:t_abund}).
Therefore, following \citet{lennonlb1}, for a given \logg the degree of veiling can be estimated by requiring that an abundance analysis results in a normal silicon abundance for the LMC. We adopt the LMC silicon abundance of $7.2\pm0.1$\,dex from \citet{hunter2009} as they methods similar to those used here, thereby minimizing systematic effects. In fact many of their sample of B-type stars are also in the NGC\,2004 cluster, and its vicinity.

\begin{figure}
    \centering
    \includegraphics[width=1.0\hsize]{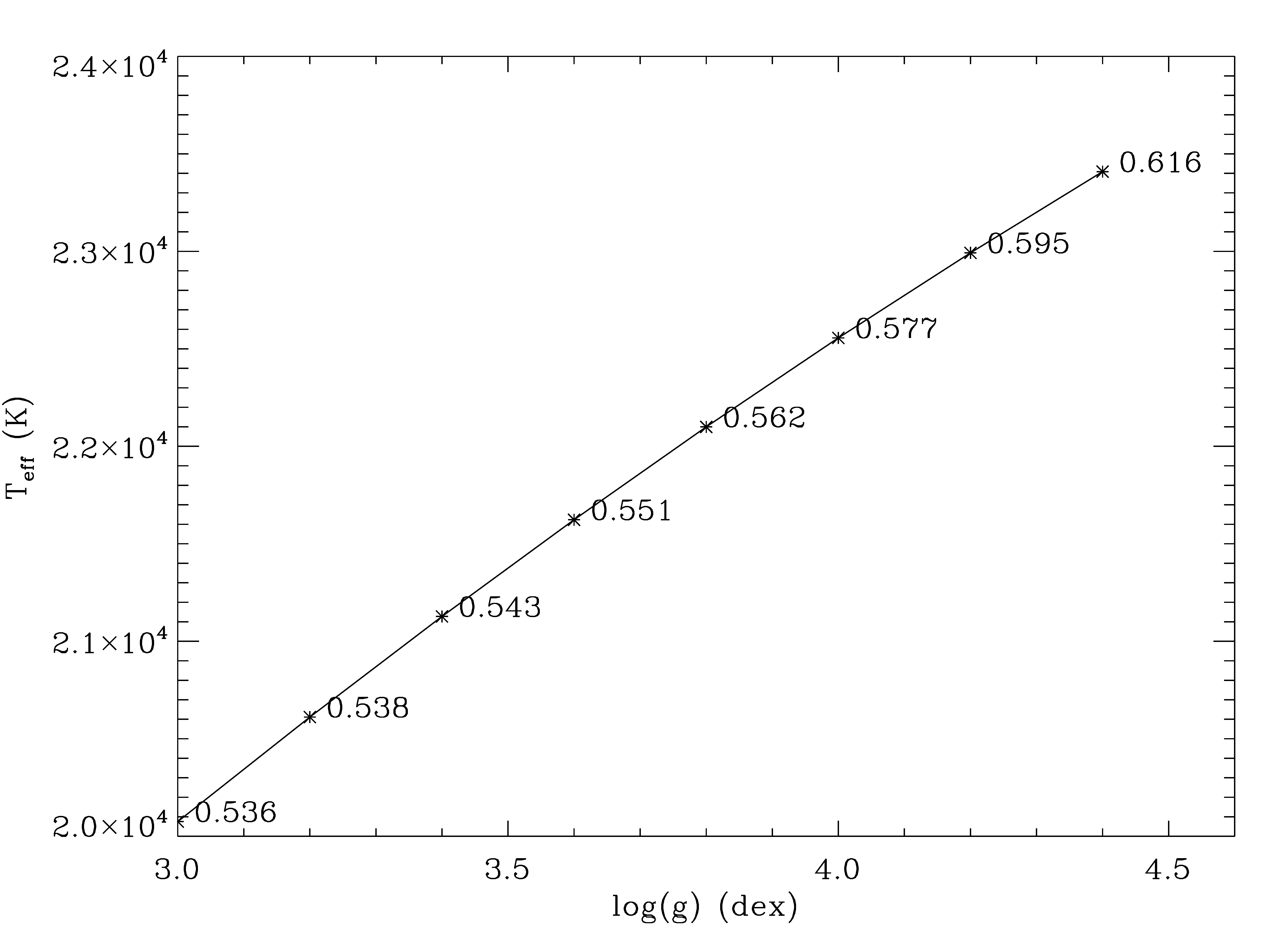}
    \caption{The locus of (\logg,$T_{eff}$) parameters, labeled with the fractional B-star contributions, that reproduce the normal LMC silicon abundance of 7.2\,dex. A helium abundance of N[He/H]=0.1 and a microturbulence of \vt=0\,\kms\ have been assumed.  }
    \label{fig:fraction}
\end{figure}

The stellar effective temperature (\teff) is relatively  well constrained by the presence of \ion{Si}{iii} lines, the near absence of \ion{Si}{ii} and \ion{Si}{IV}, and the absence of \ion{He}{ii} lines. As shown in Fig.~\ref{fig:fraction}, the contribution of the narrow-lined star to the total flux in the $V$-band lies in the range 0.54--0.62, with \teff$\sim$21500$\pm1500$\,K depending on \logg, and assuming a helium abundance by number of N[He/H]=0.1. Increasing the helium abundance to N[He/H]=0.2 decreases the flux contribution from the B-type star by $\sim$0.02--0.04 but, as we find that the range of surface compositions implies only a slight nitrogen enhancement, a helium enrichment is unlikely. Therefore we have not explored helium-rich models further. The uncertainty in [Si/H] of $\pm0.1$ translates into an uncertainty of $\mp0.05$ in the flux contribution from the B-type star.

For early type stars, the primary diagnostic of the surface gravity is the strength of pressure broadened wings of the hydrogen Balmer lines. In the present case, the composite nature of spectrum prevents use of this standard method, while a disentangling approach \citep[as in][]{Shenar_2020} is also not feasible since, for example, there are effectively only two epochs of the H$\gamma$ observations. 
However, the Balmer lines, or the gravity sensitive Diffuse \ion{He}{I} lines, offer an alternative method for determining the degree of veiling, and by inference \logg, as their sharp Doppler cores provide a method for assessing which of the models indicated by Fig.\,\ref{fig:fraction} best removes that feature from the residual spectrum. The Balmer lines are not used since modeling the line cores themselves depends on the mass-loss rate \citep{najarro1996}, which is not included in our models. By contrast, the weaker \ion{He}{I} lines are well modeled by NLTE plane parallel hydro-static equilibrium codes \citep[see for example][]{irrgang}.

Therefore, as the helium abundance is essentially solar (from above), we use the \ion{He}{I} 4471\,\AA\ line to constrain \logg. This is the strongest Diffuse line in our data, it has the best understood line broadening, and lies in the region of highest s/n of the VLT-FLAMES data.  Furthermore, from examination of FSMS data for other stars within $\sim$30\arcsec\ of NGC\,2004\#115, ELS numbers 23, 73, 98 and 118, it is also clear that nebular emission is insignificant. For this test we merged the subexposures within each observing block of the VLT-FLAMES data, with appropriate velocity corrections. The results, indicated in Fig.\,\ref{fig:logg}, demonstrate that the residuals of the subtracted Doppler core imply a \logg$\ge$3.75, though the upper bound is poorly constrained. We therefore adopt \logg$\simeq 4.0$, and hence \teff=22600\,K and a B-type star flux contribution of 0.58. We recognize that \logg\ is uncertain, however, as discussed in section 4, the star's luminosity, as derived below, provides a more precise evolutionary mass, which is consistent with \logg$\simeq 4.0$ (the determination of the radius and luminosity is insensitive to the value of \logg). 

Using the above information, and comparing the observed de-reddened $V$-band flux (subsect.~3.4) with that computed using the filter function of \citet{bessell1990}, and assuming a distance to the LMC of 49.59\,kpc \citep{pietrzynski2019}, leads to  estimates of the stellar radius and luminosity of the B-type star as indicated in Table \ref{table:t_abund}. This table also indicates the derived chemical composition and, for comparison, equivalent results for the case of zero veiling. The composition of the star is very similar to baseline LMC abundances, with perhaps a modest nitrogen enhancement being apparent. 
As discussed in Appendix~\ref{appendix:atmosphere}, while we adopt a best fit microturbulence velocity (\vt) of \vt=0\,\kms, increasing this to \vt=2\,\kms\ would strengthen the theoretical metal lines and lead to a decrease in the B-star contribution to $\sim$0.51, for the above value of \logg.  

\subsection{The properties of the broad-lined star}

%\textcolor{red}{To be completed with input from PLD}

\begin{figure}
    \centering
    \includegraphics[width=\hsize]{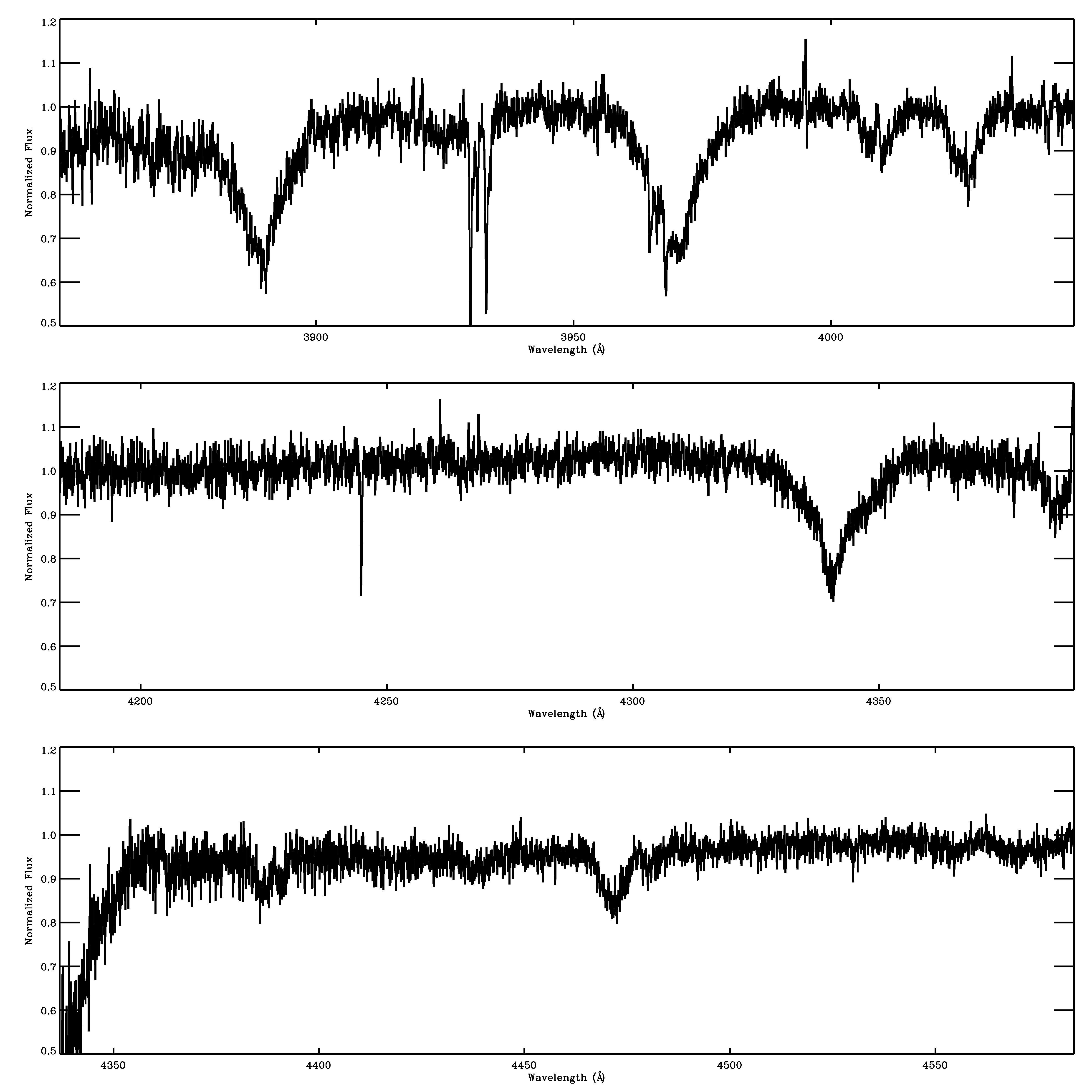}
    \caption{Subtracted spectra of the proposed tertiary for the HR02, HR04 and HR05 regions.
    We have not included HR03 due to its low signal-to-noise (discussed in the text), or HR06 as it is relatively featureless. Close examination of this spectrum may show reveal the presence of over-subtracted metal lines, but these are artifacts due to the use of a specific model spectrum from our grid with mean surface abundances a little different from the final abundance determination of the primary. Spectra were approximately normalized using a low order polynomial fitted to continuum regions.}
    \label{spectra:tertiary}
\end{figure}

\begin{figure}
	\centering
	\includegraphics[width=1.0\hsize]{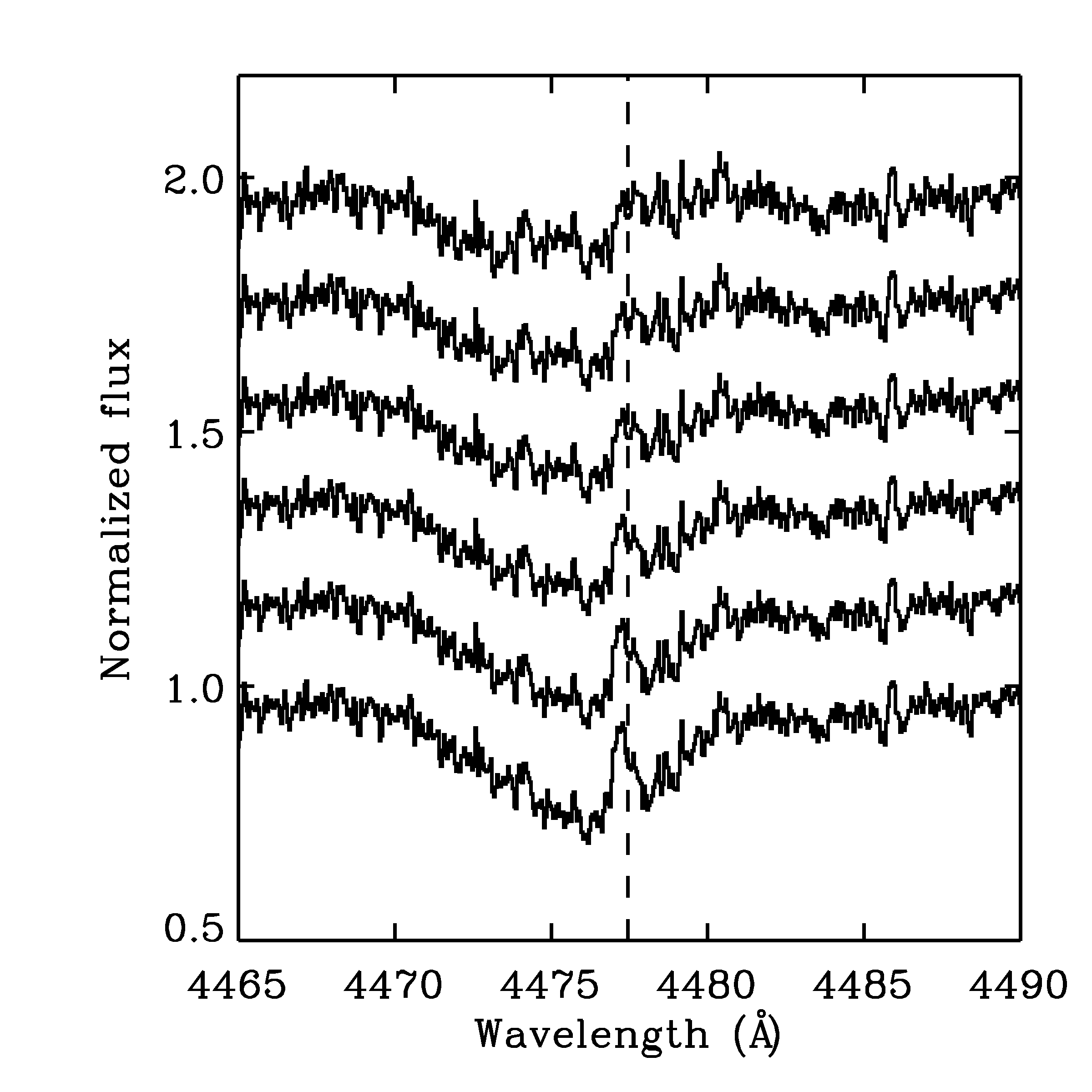}
	\caption{Illustration of the residual spectra obtained by subtracting theoretical spectra scaled by the fractional contribution from the HR06 OB2 data in the vicinity of \ion{He}{i}\,4471\,\AA\ line. The value of \logg\ increases in steps if 0.25\,dex from \logg=3.00\,dex for the lower curve to 4.25\,dex for the uppermost curve, spectra are offset by 0.2 normalized flux units. \teff\ values follow the relationship shown in Fig.\,\ref{fig:fraction}. The emission feature at the position of the \ion{He}{i} line core (dashed line), caused by over-subtraction of the Doppler core in the model, becomes weak or absent for \logg$>3.75$. }
	\label{fig:logg}
\end{figure}

Adopting the previously determined parameters for the primary of \teff=22600\,K, 0.58 fractional contribution to the V-band flux, helium abundance N[He/H]=0.1, and \logg=4.0\,dex, we subtract the model spectrum from the data, appropriately weighted and Doppler shifted, to produce a residual spectrum. To improve the signal-to-noise we merge individual exposures per OB of the VLT-FLAMES data.  The resultant spectra exhibit \ion{He}{I} lines that support our assumption that the tertiary is a mid to early B-type star, see Fig.\,\ref{spectra:tertiary}, moreover we can estimate its radial velocity from the \ion{He}{I} and Balmer lines (noting that the presence of emission may impact the results).

In determining radial velocities and \vsini\ estimates we used the \ion{He}{I} lines at 4009/4026\,\AA\ (both in HR02 spectra) and 4387/4471\,\AA\ (both in HR05 spectra), other lines were found to be too weak in the residual spectrum to provide useful measurements or a confident detection.  We followed the methodology of \citet{duftonbe}, who determined radial velocities, and \vsini\ estimates, for a large sample of Be stars in the LMC and SMC using the \ion{He}{I} lines. 
Briefly, \vsini\ was estimated for each line using two different methodologies, profile fitting (PF) and Fourier transform (FT), and are listed in Table\,\ref{table:broadvsini}. The overall means for all lines agree quite well, but there are some differences between the PF and FT estimates per line and between lines; we adopt \vsini$\sim$300\,\kms. Radial velocities were determined from the centroids of the lines resulting from the PF approach. The laboratory wavelengths for the allowed components of the \ion{He}{i} 4026 and 4471\,\AA\ lines were corrected for the presence of forbidden components in their blue wings, based on simulations for an extensive grid of rotationally broadened model spectra. The radial velocity estimates are listed in Table\,\ref{table:broadrv} and, while formal errors are not given, there are differences of up to 30\,\kms\ between lines in a given VLT-FLAMES observation. This probably reflects the difficulty in measuring line centroids of the \ion{He}{i} lines in such a broad lined star, but may also be partly due to a weak emission component (presumably from the disk or faint diffuse nebulosity) that is distorting the line profiles in the VLT-FLAMES data.

\begin{table}
	\caption{Measured projected rotational velocities (\vsini) of the broad-lined tertiary as determined from the residual spectrum after subtraction of the model for the primary. Results are given for the profile fitting (PF) and Fourier transform (FT) methods. Mean values, with their standard deviations, are given in the last row.}
	\label{table:broadvsini}
	\centering
	\begin{tabular}{llcc}
	\hline\hline
	Observation & Line &  \multicolumn{2}{c}{\vsini\,(\kms)} \\
	 & & PF & FT \\ \hline
	HR02 OB1 & He\,\sc{I} 4009 & 307 & 337 \\
	HR02 OB1 & He\,\sc{I} 4026 & 285 & 296 \\
	HR02 OB2 & He\,\sc{I} 4009 & 319 & 306 \\
	HR02 OB2 & He\,\sc{I} 4026 & 326 & 300 \\
	HR05 OB1 & He\,\sc{I} 4387 & 347 & 326 \\
	HR05 OB1 & He\,\sc{I} 4471 & 280 & 291 \\
	HR05 OB2 & He\,\sc{I} 4387 & 303 & 327 \\
	HR05 OB2 & He\,\sc{I} 4471 & 279 & 289 \\
	 & & 306$\pm$24 & 309$\pm$18 \\ \hline
    \end{tabular}
\end{table}

\begin{figure}
    \centering
    \includegraphics[width=1.0\hsize]{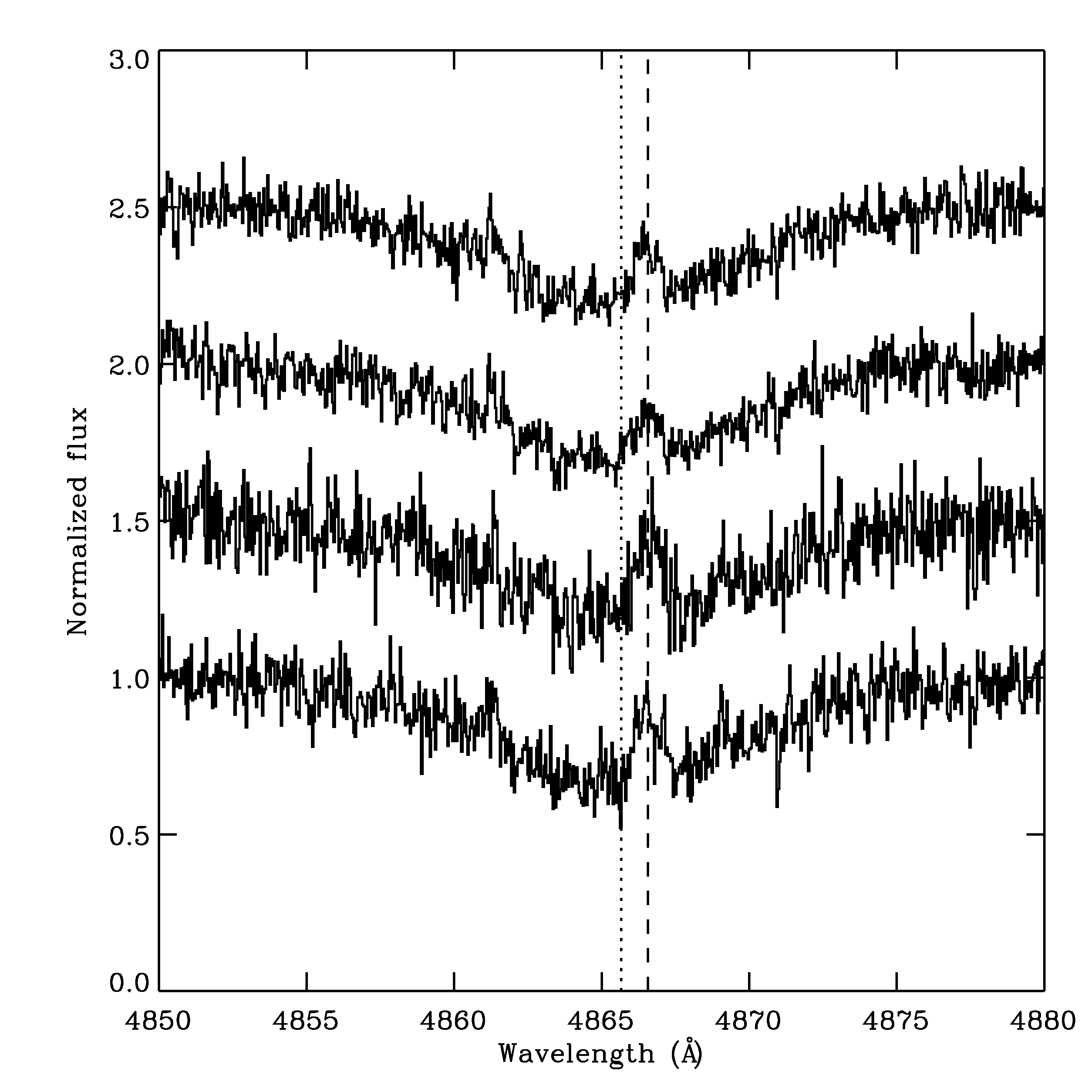}
    \caption{Residual (i.e., primary subtracted) H$\beta$ profiles from the SALT/HRS data, observations 1 through 4 are from bottom to top, shifted by 0.5 units for clarity. The vertical lines indicate the mean positions of the absorption components (dotted line) and emission components (dashed line), from Tables \ref{table:broadrv} and \ref{table:emrv}.}
    \label{fig:hbeta}
\end{figure}

We also measured the radial velocities of the residual H$\gamma$ profiles from the VLT-FLAMES data (two epochs of HR04 data), and the H$\beta$ profiles from the SALT-HRS data (all four epochs), though for these we fit rotationally broadened model profiles to the measurements, also listed in Table\,\ref{table:broadrv}. The H$\beta$ line is particularly well modeled as we can explicitly account for the weak and narrow emission component in fitting the residual spectrum (next subsection, and Fig.~\ref{fig:hbeta}).
These results in particular imply an approximate mean velocity of +267\,\kms, compared to the systemic velocity of the binary of +320\,\kms\ at this epoch, and show very little evidence of variability despite spanning six days and very different phases of the inner binary.  Evidently the residual broad-lined spectrum is that of the tertiary, and is not a component of the short period inner binary.
For convenience the broad absorption line velocities are also plotted in Fig.~\ref{fig:solutions}.

\begin{table}
	\caption{Measured radial velocities ($v_r$) of the broad-lined tertiary as determined from the residual spectrum after subtraction of the model for the primary.}
	\label{table:broadrv}
	\centering
	\begin{tabular}{lllll}
	\hline\hline
	Observation & Line &  \multicolumn{2}{l}{$v_r$ (\kms)} \\ \hline
	HR02 OB1  &He\,\sc{I} 4009 & 278  \\
	HR02 OB1  &He\,\sc{I} 4026 & 311  \\
	HR02 OB2  &He\,\sc{I} 4009 & 305  \\
	HR02 OB2  &He\,\sc{I} 4026 & 275  \\
	HR05 OB1  &He\,\sc{I} 4387 & 219  \\
	HR05 OB1  &He\,\sc{I} 4471 & 205  \\
	HR05 OB2  &He\,\sc{I} 4387 & 264  \\
	HR05 OB2  &He\,\sc{I} 4471 & 236  \\
	HR04 OB1  &H$\gamma$ & 245 \\
	HR04 OB2  &H$\gamma$ & 295 \\
	SALT/HRS obs1  &H$\beta$  & 265 \\
	SALT/HRS obs2  &H$\beta$  & 269 \\
	SALT/HRS obs3  &H$\beta$  & 263 \\
	SALT/HRS obs4  &H$\beta$  & 270 \\ \hline
    \end{tabular}
\end{table}

\subsection{The emission line spectrum}

\begin{figure}
    \centering
    \includegraphics[width=1.0\hsize]{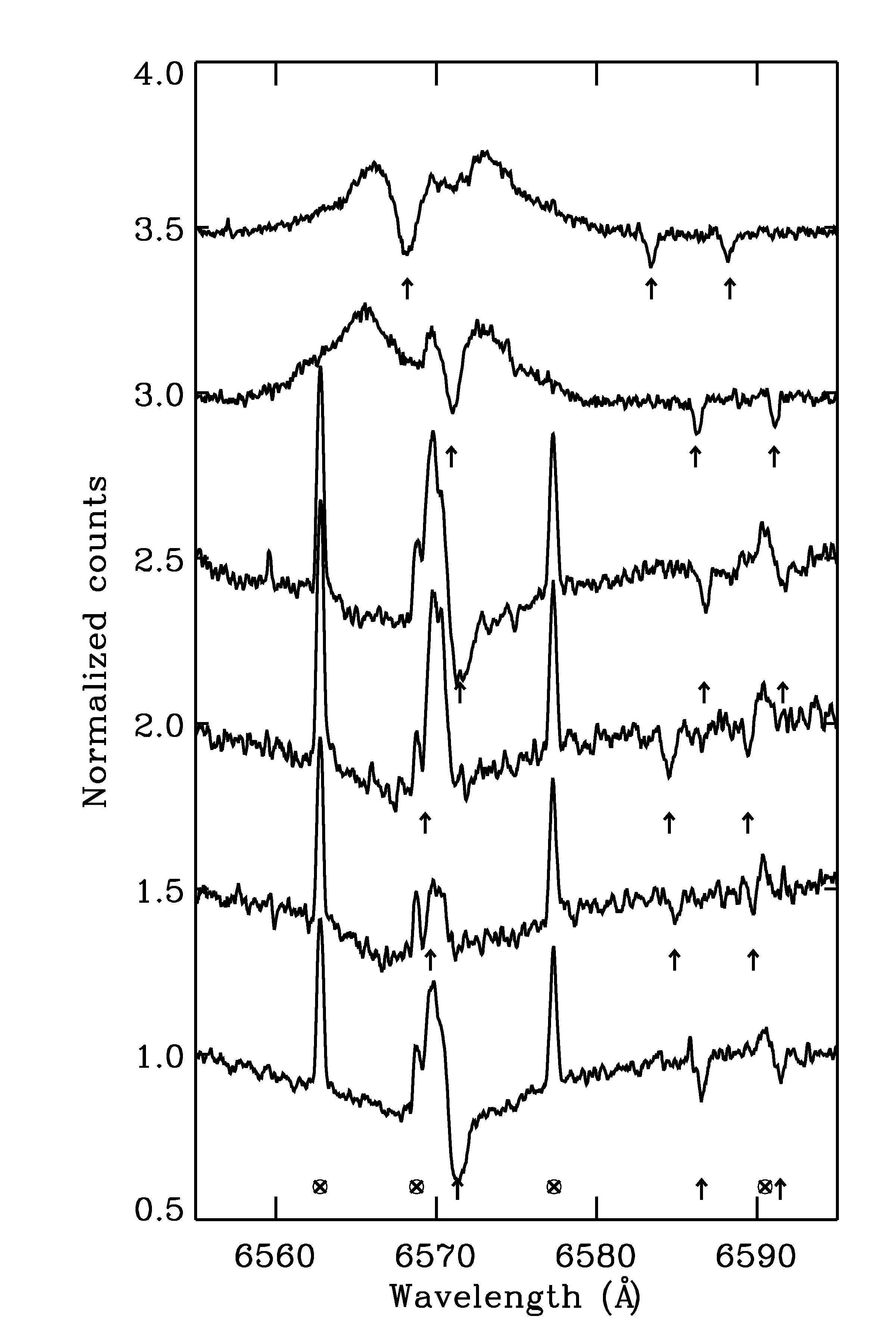}
    \caption{H$\alpha$ profile morphology in the VLT-FLAMES data (top two profiles) and SALT-HRS data (four lower profiles). Spectra have been off-set by 0.5 normalized flux units for clarity. Sharp emission features in the SALT-HRS data are night sky lines ($\otimes$). H$\alpha$ and the sky line just long-ward of 6590\,\AA\ are likely blended with weak nebular \ion{H}{i} and [\ion{N}{ii}] respectively, as discussed in the text. The position of the central dip of the H$\alpha$ Doppler core of the narrow-lined B-type star is indicated, as are the positions of its two \ion{C}{ii} lines ($\uparrow$).}
    \label{fig:all_halpha}
\end{figure}

As for our discussion of the residual spectrum above, the following analysis of the emission features utilized the merged spectra for each OB of the VLT-FLAMES data. 
Broad double-peaked H$\alpha$ emission is observed in the VLT-FLAMES data, but in the SALT-HRS data it is much weaker, narrower, and rather flat-topped, with just a hint of a double peak (see Fig.~\ref{fig:all_halpha}). Also apparent is a very weak contribution from diffuse nebular emission, best seen in the VLT-FLAMES spectra in Fig.~\ref{fig:halpha} as a weak and narrow spike in the core of the emission line (its velocity is +315$\pm$5\,\kms). Diffuse nebular emission is observed at this velocity in the sky spectra, observed with SALT-HRS at distances of 40" and 63" from the source. The strength of this emission depends on position, however, and simply subtracting the sky spectrum from the source results in over-subtracted nebular emission, and strongly distorted line H$\alpha$ line cores. Hence, the data shown in Fig.\ref{fig:all_halpha} have not been sky-subtracted and still exhibit the typical night sky emission lines observed near H$\alpha$.
The SALT-HRS data reveal that the H$\beta$ line, not observed by VLT-FLAMES, also exhibits weak emission. Very weak H$\gamma$ emission is detected in the VLT-FLAMES data, although its weakness prevented measurement of reliable emission line profiles.

\begin{figure*}
	\centering
	\includegraphics[width=0.45\hsize]{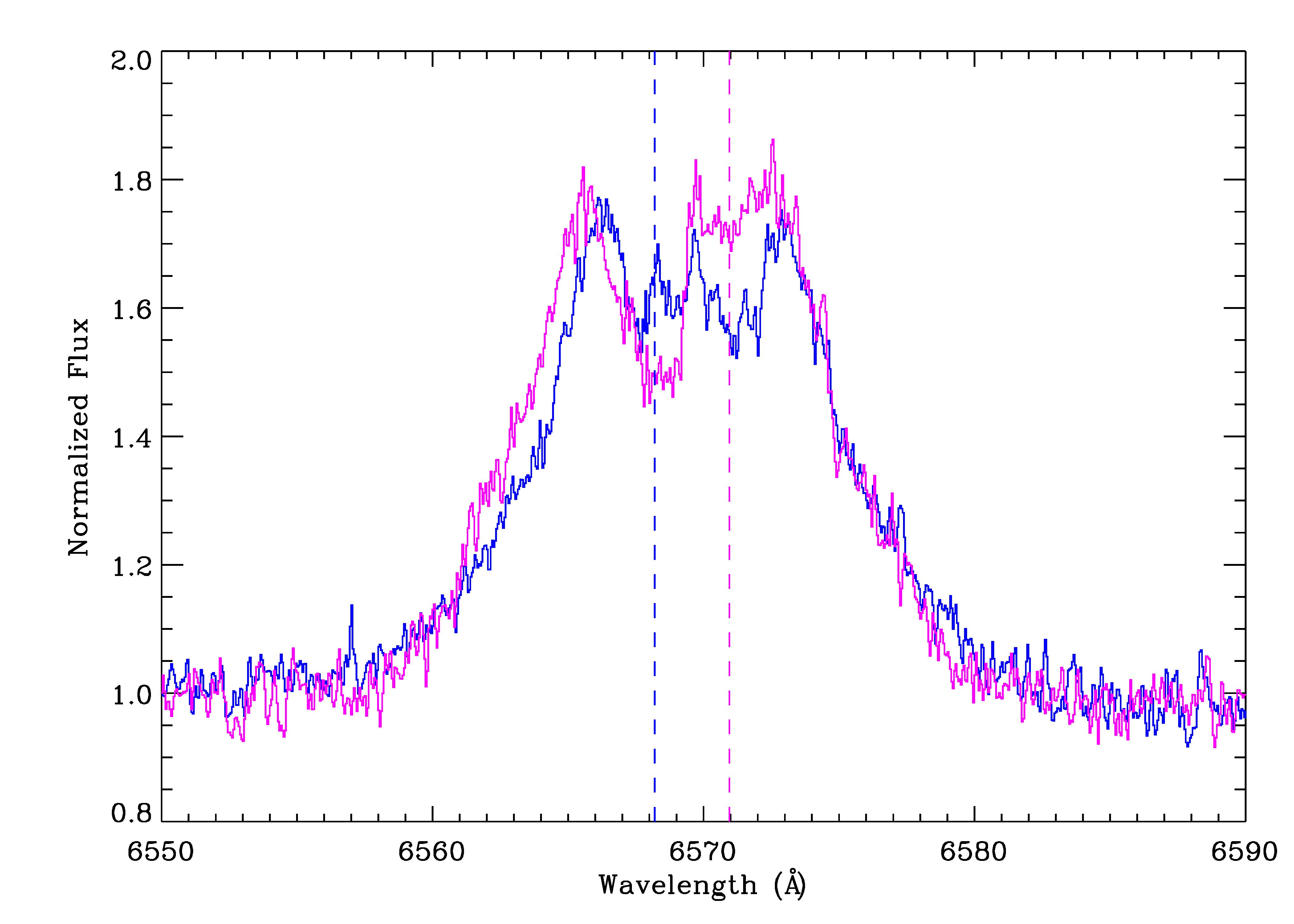}
	\includegraphics[width=0.45\hsize]{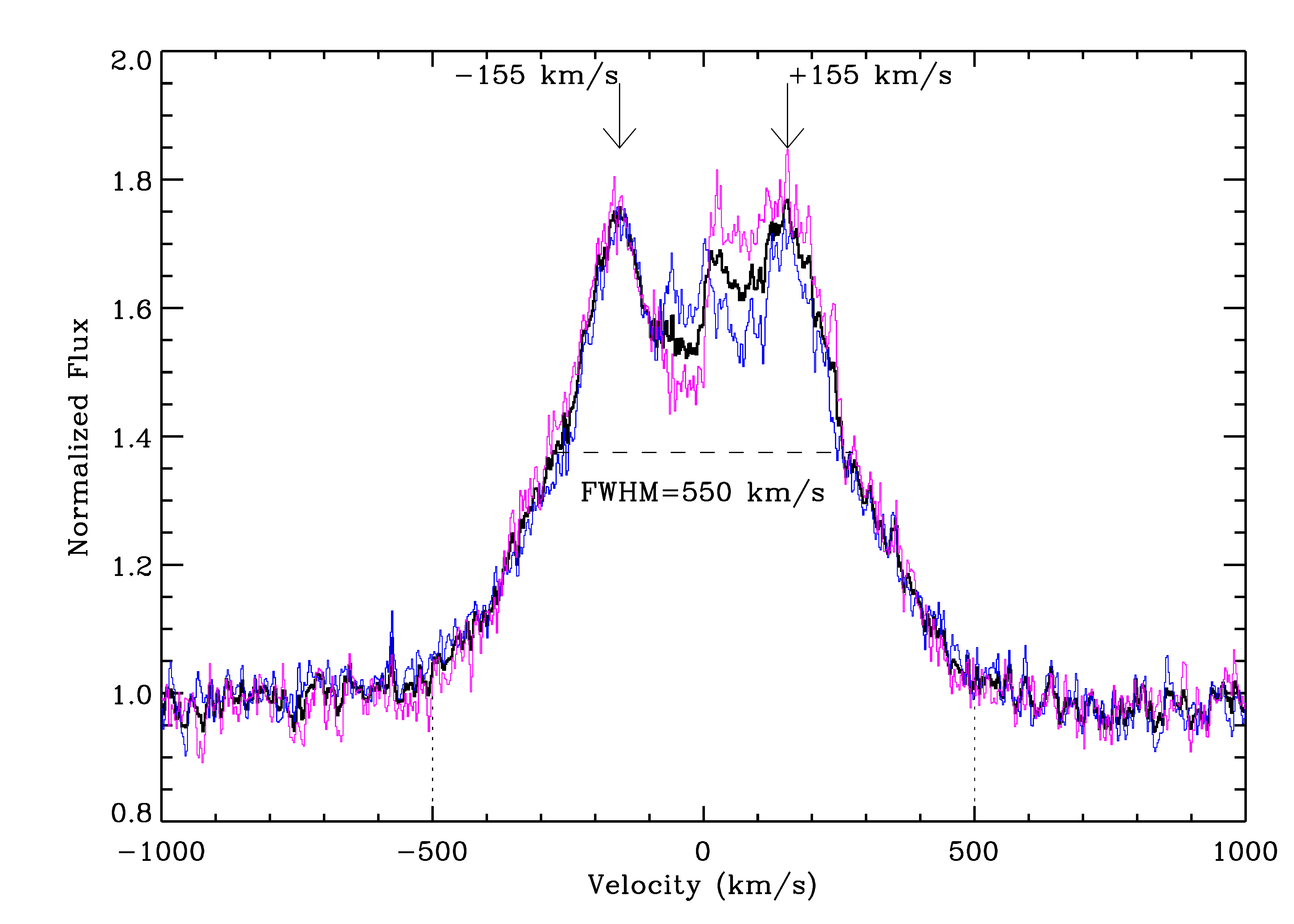}
	\caption{Left panel: VLT-FLAMES OB1 (blue) and OB2 (magenta) H$\alpha$ profiles resulting from subtraction of the underlying spectrum of the narrow-lined B-type star. Horizontal lines, same color coding, represent the rest frame velocity of the primary at these epochs. Note that a small residual velocity shift is still apparent (cf Fig.\,\ref{fig:halpha}). Right panel: the same profiles plotted in their rest frame velocities; OB1 and OB2 spectra in blue and magenta respectively, the merged spectrum is plotted in black. Also illustrated are the velocities of the double peaks ($\pm$155\,\kms), the full-width half-maximim (FWHM) of the emission (550\,\kms), and the extent to which emission is present in the line wings ($\pm$500\,\kms). These spectra were taken $\sim$34 days apart and some variability is apparent within the double-peaked structure of the profile.}
	\label{fig:halpha_velocity}
\end{figure*}

There appear to be small velocity shifts in the broad H$\alpha$ emission, as shown in Fig.~\ref{fig:halpha}, that are anti-correlated with those of the primary. 
However, as this apparent motion may be due to the neglect of the underlying absorption from the B-type star that distorts the emission profile \citep{abdulmasih,elbadry2020}, we subtracted this contribution from the data, shown in
Fig.~\ref{fig:halpha_velocity}. A small shift between the profiles is still observed and, given that to a good approximation they are symmetric outside of the central dip, we determine their radial velocities from the position of bisector of the profiles. This results in values of 310$\pm$5 and 290$\pm$5 \kms\ for the merged OB1 and OB2 data respectively, see Table~\ref{table:emrv}.
The relative lack of motion of the emission is pronounced given the contemporaneous velocities of the B-type star of approximately +245 and +372\,\kms\ for OB1 and OB2 respectively and, while they appear to be anticorrelated, these measurements were taken 34 days apart and hence may reflect the systemic velocity drift of the inner binary. 

\begin{figure}
	\centering
	\includegraphics[width=1.0\hsize]{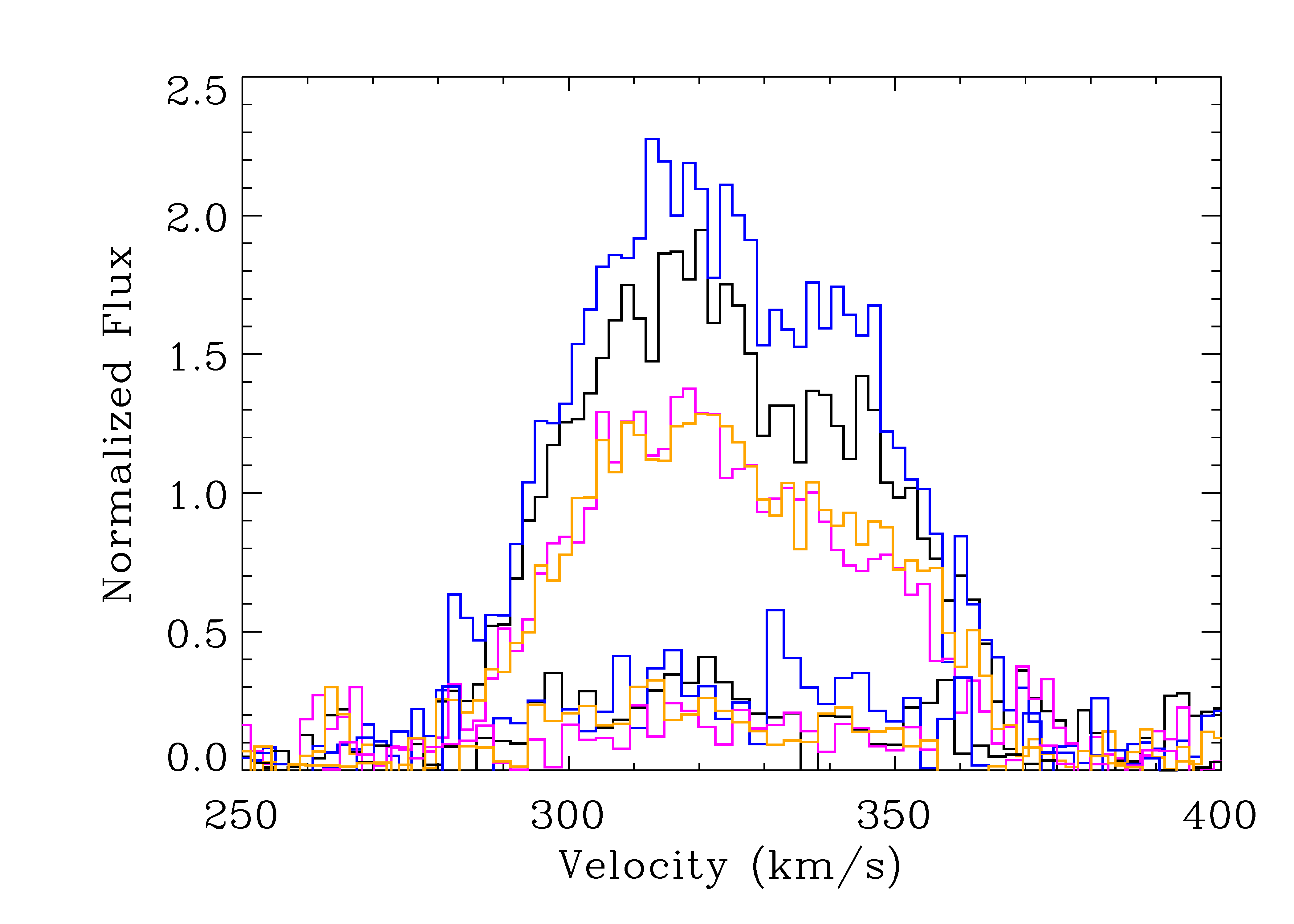}
	\caption{SALT-HRS line profile morphology in velocity space of H$\alpha$ (upper four curves) and H$\beta$ (lower four curves), color coded by observation; black -- obs1, blue -- obs2, magenta -- obs3 and orange -- obs4. The primary's underlying profiles have been subtracted from these data. While the H$\alpha$ profiles exhibit some variability of intensity their velocities are quite constant, and are well modeled with two-component Gaussian fit with velocities at 314\,\kms\ (for the ambient nebular emission) and $\sim$325\,\kms\ (for the source emission). The latter has a FWHM of $\sim$60\,\kms.}
	\label{fig:salt_vel}
\end{figure}

The H$\alpha$ emission in the SALT-HRS data is narrow, with just a hint of a double peaked structure (it as almost flat-topped) when corrected for the underlying absorption and night sky lines, see  Fig.~\ref{fig:salt_vel}.
Ambient nebular emission must also be present, for example a weak [\ion{O}{III}] 5006.84\,\AA\ line is present in the SALT-HRS on-source data and we measure its velocity to be 314$\pm$1\,\kms, with a line width of 20\,\kms\ (the resolution of the data). Very weak [\ion{N}{II}] lines are also present, with a consistent velocity. We also see weak Balmer emission with this velocity and width in the sky spectra, though as noted the sky fibers were quite far from the source position and the line strengths variable, preventing precise sky subtraction.
The features shown in  Fig.~\ref{fig:salt_vel} are clearly broader than those of the ambient nebular emission, hinting at an additional emission component at higher velocity. We therefore fitted the H$\alpha$ lines with a two-component Gaussian model, fixing position and width of one Gaussian at the values measured for the ambient emission, and optimizing the fit for its peak height and the parameters for the second component. 
The results for the velocity of this second component, which we identify with the source, are also summarized in Table~\ref{table:emrv}, as are the H$\beta$ results, for which we used a single component fit as the ambient feature there should be much weaker than in H$\alpha$. Clearly the unknown intensity of the ambient nebular emission is an issue, that could be resolved with high resolution spectroscopy. Nevertheless, the morphology of the emission is consistent with the source emission being at the higher velocity as the velocity of the ambient H$\alpha$ emission is well constrained.
From Table~\ref{table:emrv} we see that the H$\alpha$ velocities, taken over a few days and at differing phases, are constant to within a few \kms. The weaker H$\beta$ line, while it exhibits a hint of anticorrelation with the narrow lined star, has velocities consistent within the errors of the measurements. However, as already noted, the emission line velocities at this epoch are very different from the radial velocities of the broad-lined star (Table~\ref{table:broadrv}), and quite similar to the systemic velocity of the inner binary. On this evidence therefore we tentatively associate the source emission with the inner binary, and not with the broad-lined tertiary, although we discuss alternative scenarios in Sect.~5.

\begin{table}
	\caption{Radial velocities of H$\alpha$ and H$\beta$ emission lines. For comparison the radial velocities of the narrow-lined B-type star are also entered.}             
	% title of Table
	\label{table:emrv}      % is used to refer this table in the text
	\centering                          % used for centering table
	\begin{tabular}{l l l l}        % centered columns (4 columns)
		\hline\hline                 % inserts double horizontal lines
		 & Emission & Emission & Stellar \\
		Observation & Line & velocity & velocity \\
		\hline
		HR14A OB1 & H$\alpha$ & +310$\pm$5 & +245\\
		HR14A OB2 & H$\alpha$ & +290$\pm$5 & +372\\
		SALT/HRS obs1  & H$\alpha$ & +326$\pm$1 & +397\\
		SALT/HRS obs2  & H$\alpha$ & +324$\pm$1 & +298\\
		SALT/HRS obs3  & H$\alpha$ & +324$\pm$1 & +313 \\
		SALT/HRS obs4  & H$\alpha$ & +325$\pm$1 & +390\\
		SALT/HRS obs1  & H$\beta$ & +320$\pm$3 & +397\\
        SALT/HRS obs2  & H$\beta$ & +326$\pm$3 & +298\\
        SALT/HRS obs3  & H$\beta$ & +321$\pm$3 & +313 \\
        SALT/HRS obs4  & H$\beta$ & +318$\pm$4 & +390\\
		\hline                                                   % inserts single horizontal line 
		\end{tabular}
\end{table}

\begin{figure}
   \centering
   \includegraphics[width=\hsize]{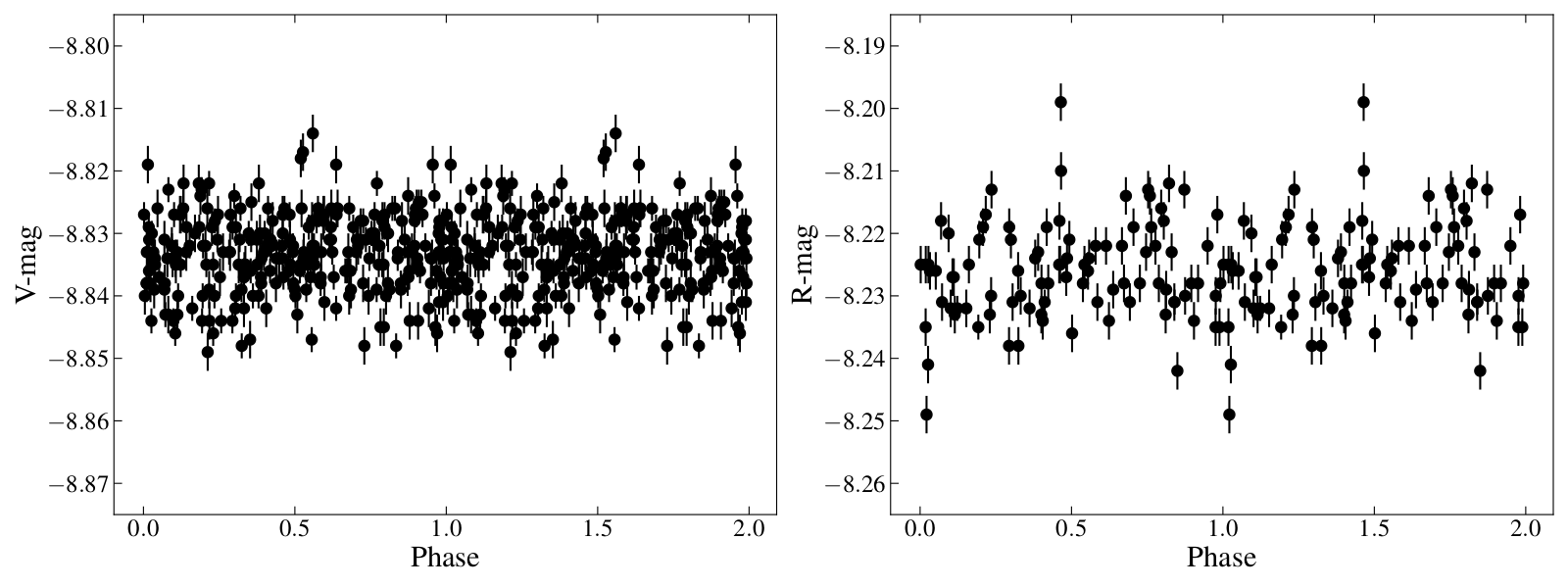}
      \caption{MACHO $V$ and $R$ photometry spanning MJD\,49001.6091 to 51542.5763, folded to the 2.92\,d period of the system, illustrating the lack of variability in both bands. The magnitudes are instrumental; zero points and transformations to Kron-Cousins $V$ and $R$ magnitudes can be found in \citet{alcock1999}.}
         \label{fig:macho}
   \end{figure}

\subsection{Photometry, optical light curve, and X-ray flux limit}

The available optical and infrared (IR) data are described in Appendix \ref{appendix:sed}, and in general are in good agreement with the results of the atmospheric analysis, though there is evidence of a small IR excess. 
The magnitude of the source has been extremely constant, illustrated by approximately 7 years of phase-folded MACHO data in Fig.~\ref{fig:macho}, that limits possible ellipsoidal light curve variations to less than 0.005$^m$ in $V$ and 0.007$^m$ in $R$ in amplitude, or around 0.5--0.6\%.  The lack of variability of the system is also reflected in the $Gaia$ data with the EDR3 $G$-band flux indicating an uncertainty of 0.1\%, and simulating the modulation of ellipsoidal light variations in the data we estimate the amplitude of these variations to be less than 0.3\%, see Appendix \ref{appendix:sed} for details. 
Finally, XMM-Newton observations provide an upper limit to the X-ray flux from the system of less than 5.33x10$^{33}$ ergs\,s$^{-1}$ (about 1.4\,$L_\odot$) in the 0.2--12\,keV band. 

It the above context it is interesting that classical Be stars are well known variables, \citet{hubert} find that 86\% of early Be stars are subject to rapid variability of at least 0.02$^m$, equivalent to about 0.01$^m$ given the veiling effect here, while NGC2004\#115 is constant to better than 0.002$^m$ in the Gaia $G$ band. In addition its IR excess is very small for a typical Be star, for example $J-[3.6]=-0.13^m$ compared with a typical value of $-0.6^m$ \citep{bonanoslmc,bonanossmc}. 

\section{Discussion}

In this section we present some evolutionary scenarios and discuss to what extent they can explain the primary constraints on the system. We make the simplifying assumption that the orbital dynamics of the inner binary are not significantly perturbed by the tertiary, and adopt the mass function for this system from Table~\ref{table:orbit}. Other hard constraints on this system are the ellipsoidal light variations ($f_{\rm el}$) of less than 0.6--0.3\%; the radius and luminosity of the primary from Table \ref{table:t_abund}, its low \vsini\ $<$15\,\kms, and the lack of any eclipse that defines the maximum possible sin$i$. 
The value of \vsini, together with the period and stellar radius, then define the sin$i$ required to ensure synchronous rotation of the B-type star, or sin$i$<0.15. Synchronous rotation would also imply that the equatorial rotational velocity of the B-type star is 97\,\kms. 

The mass of the primary (or m$_1$) is a critical parameter of the system that is not well constrained from observations (subsection 3.1).
The surface gravity of the primary implies a spectroscopic mass ($M_{\rm sp}$) of $\sim$9\,$M_\odot$, though with a large uncertainty giving a mass range of 5--16\,$M_\odot$. However, the extremes of this mass range can be ruled out as being inconsistent with the star's luminosity, for example, a 16\,$M_\odot$ star would have a luminosity of 4.7--4.8\,$L_\odot$, well above the estimate derived from its radius and \teff. 
In fact we can also define an evolutionary mass ($M_{\rm ev}$) using {\sc bonnsai}\footnote{BONNSAI is available at www.astro.uni-bonn.de/stars/bonnsai} \citep{schneider2014,schneider2017} that is based on the models of \citet{brott} and \citet{kohler}, which were calibrated using B-type stars observed by the FSMS, many of which are in NGC\,2004 \citep{hunter2009}. 
As input we use \teff=22600\,K, R$_1$=5.6\,R$_\odot$, and \vsini\ as above, resulting in estimate of $M_{\rm ev}$=8.6\,$M_\odot$ and an age of 22\,Myrs. 
This is in reasonable agreement with the spectoscopic mass above, though better constrained, and consistent with the age of NGC\,2004. Note that if this B-type star is a product of binary evolution then these mass and age estimates may not be applicable. 

Alternatively the narrow lined B-type star might be a low mass stripped helium star \citep{wellstein,goetberg}, with a more massive ($\sim$10\,$M_\odot$) Be-star companion, as has been proposed for LB-1 and HR6819 \citep{shenar,bodensteiner2020,elbadry2021}. In this picture, the stripped star is in a short-lived ($\sim$10$^5$\,yr) transition as it evolves rapidly to its stable helium burning phase. Interpolating between the stripped star evolutionary tracks illustrated in the HR-diagram (HRD) in Fig.~3 of \citet{langergamma} we find that the luminosity of the narrow-lined B-type star of 3.87\,$L_\odot$ corresponds to a stripped star mass of approximately 1.4\,$M_\odot$, and with \logg$\sim$3.2.

\subsection{The narrow-lined B-type star as a stripped star}

\begin{figure}
	\centering
	\includegraphics[width=\hsize]{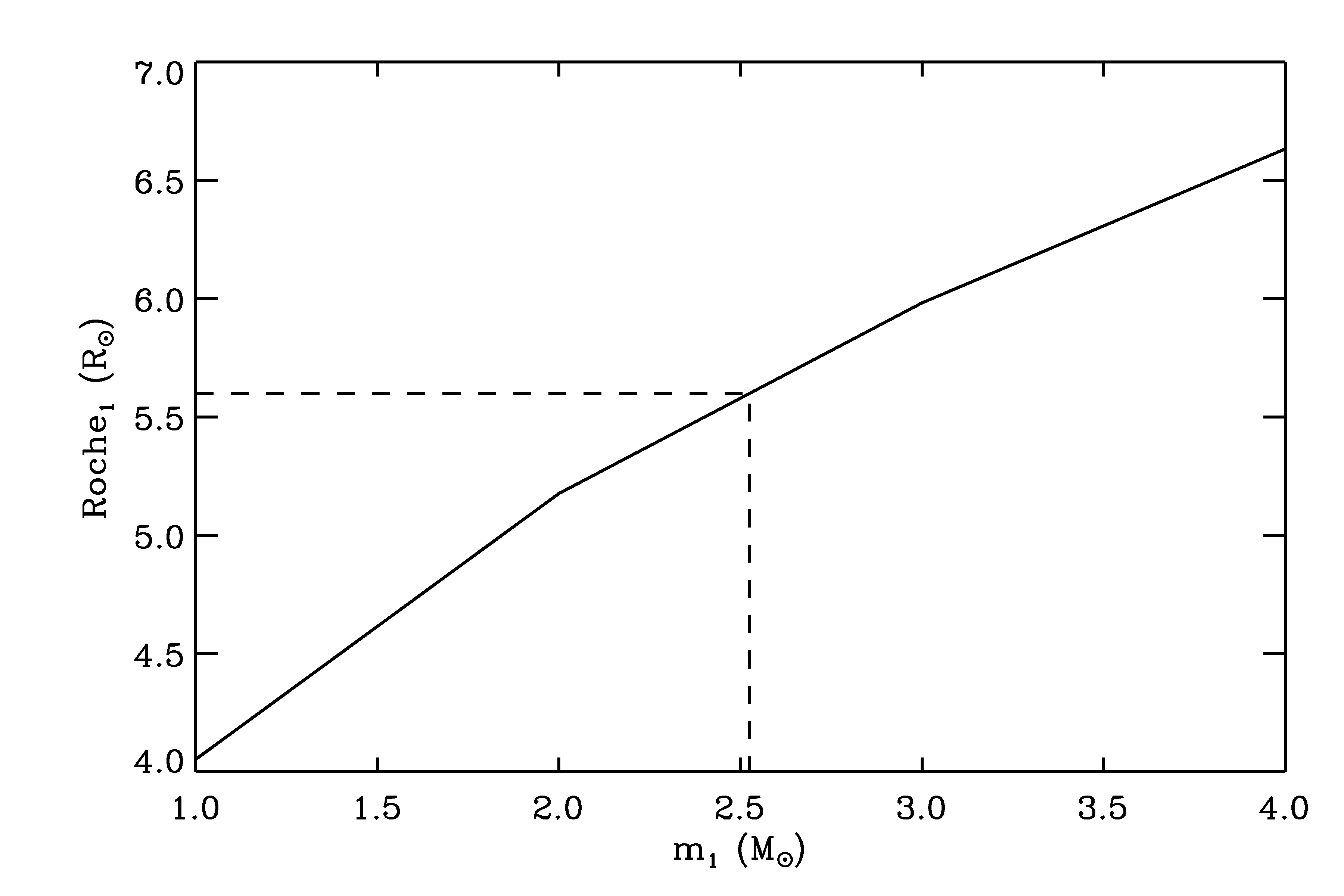}
	\caption{Plot of the mean value of the Roche radius of the primary versus primary mass for the binary parameters relevant here. The area within the dashed lines (lower left) is prohibited due to the primary's Roche radius shrinking to less than the size of the star. We plot the mean value of the Roche radius in this figure, however for all values of sin$i$ greater than 0.1 the actual value is only a weak function of sin$i$.}
	\label{fig:stripped}
\end{figure}

Considering first the stripped star scenario with m$_1$=1.4\,$M_\odot$ we find that the Roche radius is smaller than the stellar radius by approximately 20\%, irrespective of the value of sin$i$. The inflated stripped star is just too big to fit into such a short period system. The situation is illustrated in Fig.~\ref{fig:stripped} which demonstrates that for primary masses less than approximately 2.5\,$_\odot$ the Roche radius is simply too small for the size of the star. 
We therefore conclude that the observations are inconsistent with a stripped star nature for the primary.

\subsection{The primary as a massive star}

\begin{figure}
	\centering
	\includegraphics[width=\hsize]{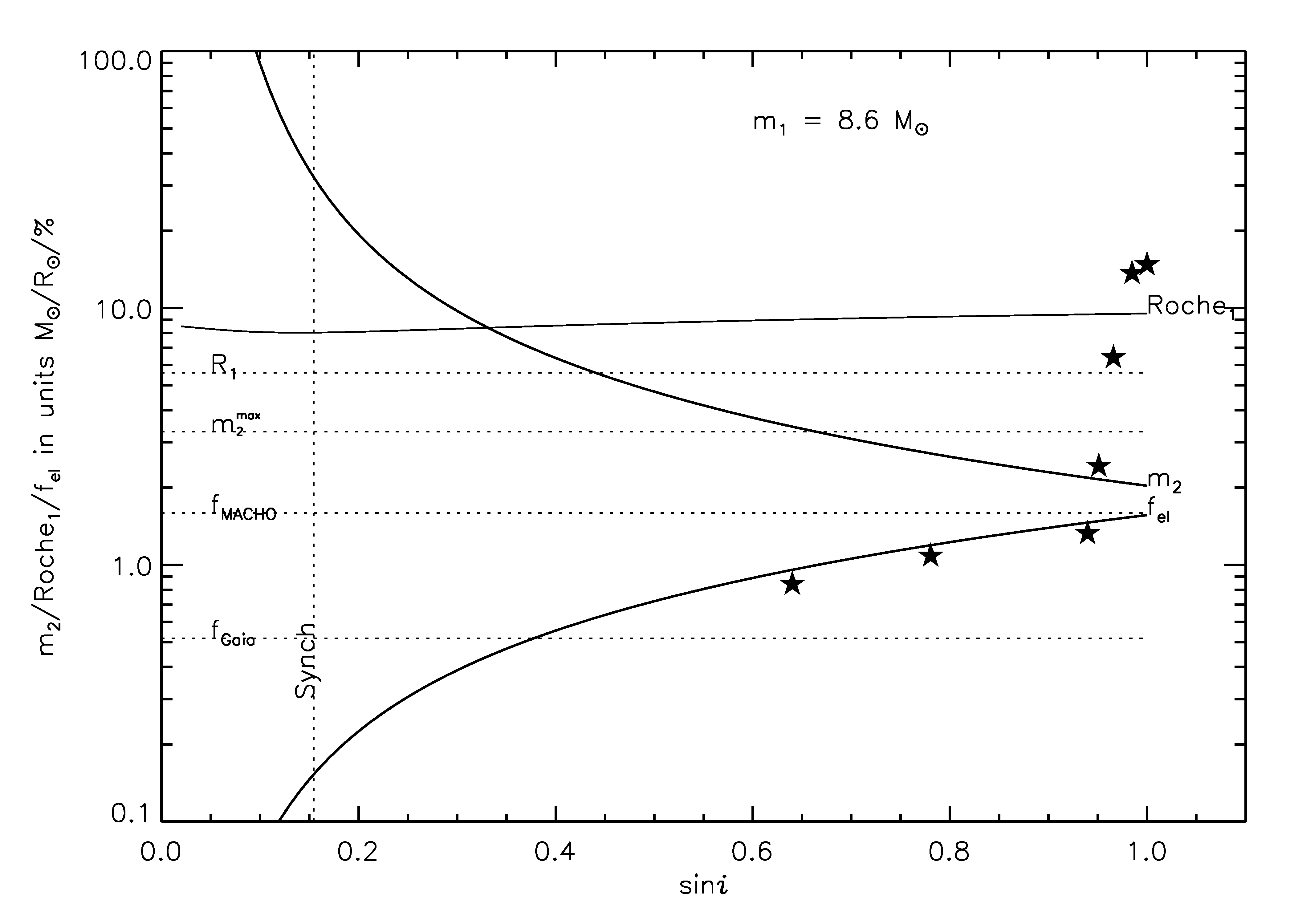}
	\caption{Diagnostic diagram assuming a main sequence primary with a mass of 8.6\,$M_\odot$. The solid curves represent, as a function of sin$i$, the secondary mass (labeled m$_2$), the Roche radius of the primary (Roche$_1$), and the maximum percentage amplitude of ellipsoidal light variations expected ($f_{el}$) of the primary. Filled stars ($\star$) illustrate the percentage amplitude of ellipsoidal light variations predicted by {\tt nightfall} assuming a binary consisting of the proposed primary and a main sequence secondary of appropriate mass and radius.  The increase at high sin$i$ indicates the presence of an eclipse. Dotted horizontal lines illustrate the constraints implied by the observations, the stellar radius (R$_1$), the limits on ellipsoidal light variations (f$_{MACHO}$ and f$_{Gaia}$), and the maximum mass that is consistent with the absence of secondary stellar spectrum in the data. The vertical line (labeled Synch) indicates the upper limit on sin$i$ implied by synchronous rotation and our measured upper limit on \vsini\ of 15\,\kms. 
	}
	\label{fig:meq8}
\end{figure}

Adopting m$_1= M_{\rm ev}=$8.6\,$M_\odot$ for the primary we construct the diagnostic diagram in Fig.~\ref{fig:meq8}. Here we plot m$_2$ (the secondary mass), Roche$_1$ (the Roche radius of m$_1$) and the predicted $f_{\rm el}$ as a function of sin$i$. Also indicated, on the vertical axis, is our estimate of the radius of the primary, R$_1=5.6\,R_\odot$, that fits comfortably inside its Roche radius, which is roughly twice the stellar radius, for all reasonable values of sin$i$.
We use the analytic approximation of \citet{Morris85} for $f_{\rm el}$ and ignore the brightness of the secondary, that is, we assume that only the primary is bright. We also used the binary simulation package {\tt nightfall}\footnote{https://www1.physik.uni-hamburg.de/en/hs/group-schmidt/members/wichmann-rainer/nightfall.html} to model binary systems for specific main sequence secondary stars (see for example Table~\ref{tab:C.1}), at appropriate values of sin$i$. From the resulting light curves we also plot the maximum percentage variation we would expect to see in the data. Note that at high sin$i$ values the light curves should display eclipses, that are not seen in the data. The observed limits on $f_{\rm el}$, labeled $f_{\rm MACHO}$ and $f_{\rm Gaia}$ for the MACHO and Gaia data respectively, are also indicated. The latter is particularly constraining, implying sin$i$<0.4.

The upper limit on sin$i$, implied by synchronous rotation, is also indicated in Fig.~\ref{fig:meq8}, and moreover implies that sin$i<0.15$. Hence the secondary should have a mass of at least $m_2\sim 25\,M_\odot$, and should be a BH since we would easily see such a massive star.  Normal stars in a binary of this period have circularization ($\tau_{circ}$) and synchronization ($\tau_{synch}$) timescales on the order of 10$^6$ and 10$^4$--10$^5$ years respectively \citep{hurley2002}, and for this specific case we estimate $\tau_{synch}$=0.07\,Myr. We are led to the conclusion that the B-type star is likely in synchronous rotation around a stellar mass BH.

We add one additional constraint on the vertical axis, namely the mass (m$_2^{max}$) above which a narrow-lined main sequence secondary star would be visible in the spectrum.  In Appendix \ref{appendix:secondary} we discuss in detail how we make this estimate, finding that m$_2^{max}$=3.3\,$M_\odot$. Ignoring for the moment the assumption of synchronous rotation, viable solutions for a slowly rotating secondary would imply a secondary mass range of 2.1--3.3\,$M_\odot$.
A fast rotating asynchronous secondary must be at least dimmer and less massive than the tertiary, but as is discussed in subsection 4.3, the disk emission appears incompatible with the radius of such a star.

\subsection{H$\alpha$ and IR emission}

As discussed in Sect.~3, it is unlikely that there is a classical Be star in the system as the Balmer emission appears to be associated with the inner binary, although it does not share the primary's orbital motion. We therefore assume that the emission is centered on a more massive secondary.  
The morphology of the Balmer emission lines suggests a disk, and we can therefore use velocity measurements to estimate its size.
The observational quantities from the VLT-FLAMES data from Fig.~\ref{fig:halpha_velocity} are the peak and wing velocities of 
$V_{p}=155$\,\kms\ and $V_{w}=500$\,\kms. The latter is of course a lower limit as it only measures the extent of the emitting disk visible in those data.
While the SALT-HRS emission lines also indicate a possible double peaked structure, see Fig.~\ref{fig:salt_vel}, this is difficult to measure, hence we instead use the half-width half-maximum (HWHM) velocity as a characteristic property of the disk, with $V_{h}=27.5$\,\kms.
On the assumption that these velocities reflect Keplerian motion in a disk we can infer some approximate dimensions for given values of $i$ and m$_2$ by estimating the radii implied by these velocities. Roughly speaking $V_{p}$ is related to the outer boundary radius, and $V_{w}$ to the inner boundary radius \citep{huang,horne1986}. 

\begin{figure}
	\centering
	\includegraphics[width=\hsize]{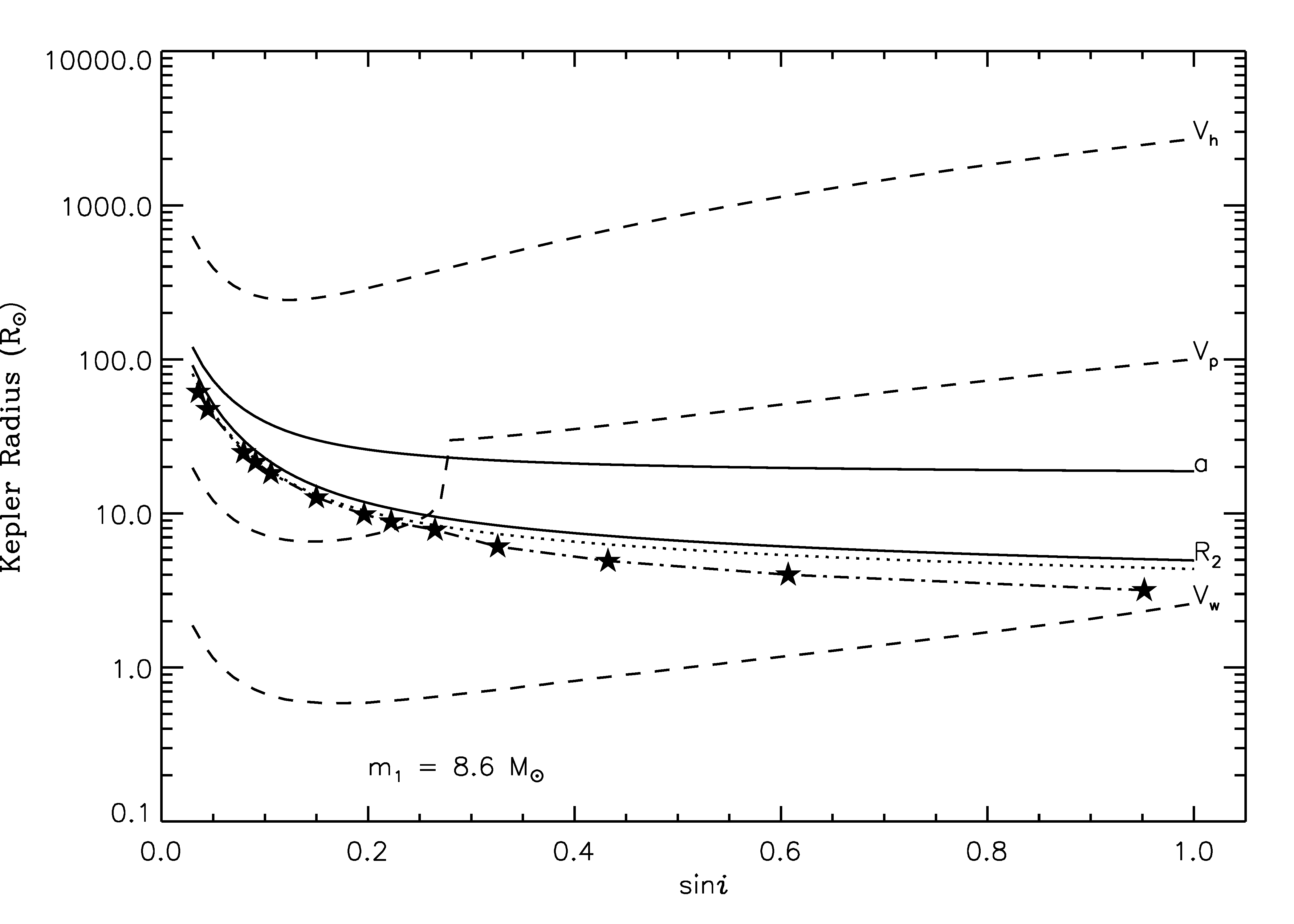}
	\caption{Plots of Keplerian radii for characteristic H$\alpha$ emission line velocities (dashed lines) as discussed in the text and labeled as follows; $V_w$ is the maximum extent of the emission in the FSMS data, $V_p$ is the velocity of that double peaked structure, and $V_h$ is the HWHM of the SALT/HRS emission. The solid curves represent the binary separation (labeled $a$) and the Roche radius of the unseen secondary (Roche$_2$). The dotted line is the limit imposed by tidal forces, while the dash-dotted line joining stars is the limiting average radius of stable orbits of the outer disk, from \cite{paczynski80}.} 
	\label{fig:diskdim}
\end{figure}

Our estimates of the radii assume that the measured velocities, corrected for sin$i$, match the Roche potential along the line joining the two masses (in the co-rotating frame,
assuming synchronous rotation), that is, in the direction of the L1 Lagrange point.  
In reality, orbits about m$_2$ in the plane of rotation are elongated along this axis but comparison with numerical solutions \cite[see for example][]{paczynski80,artymowicz} indicates this approach gives a reasonable estimate of the mean radius of stable orbits. This estimate improves with decreasing radius, though of course makes no allowance for orbit instabilities or tidal effects, that are more important at larger radii.
Fig.~\ref{fig:diskdim} illustrates the expected radii for the three characteristic velocities above, assuming a primary mass of 8.6\,$M_\odot$, compared with the binary separation and the Roche radius of the secondary. We continue to refer to m$_2$ as the secondary even though we expect m$_2$>m$_1$ in the current scenario.  We also show the locus of the secondary's Roche radius, and the binary separation ($a$), together with curves representing the maximum disk radius permitted by tidal forces \citep{artymowicz}, and maximum radius permitted by stability considerations \citep[taken from][]{paczynski80} of free-particle orbits. Similar results are found from more complex calculations for viscous disks by \citet{artymowicz}, though differences are found for nonzero eccentricity and increasing Reynolds number.

Fig.~\ref{fig:diskdim} suggests that the outer disk velocity $V_p$ implies the disk can fit inside the Roche lobe provided sin$i$<0.25. (The discontinuity in the $V_p$ curve near this point is simply an artifact, due to the velocity exceeding that at the L1 Lagrange point. Values above this point should be ignored.) $V_w$ on the other hand is well within the Roche radius for all sin$i$. The H$\alpha$ emission profile from the VLT-FLAMES data therefore suggests the presence a disk at small values of sin$i$, roughly lying between 0.7 and 7 $R_\odot$, consistent with the results from the previous subsection. 

Considering now $V_h$, from the SALT-HRS data, its value implies the gas lies at a very large distance from the binary, from a few hundred to a few thousand solar radii.  Furthermore, the variability of the disk emission confirms that the proposed disk is unstable on timescales of years. 
We have no explanation for this other than to suggest the inner binary has somehow ejected its disk, possibly as a result of a close encounter with the tertiary.

\begin{figure}
    \centering
    \includegraphics[width=\hsize]{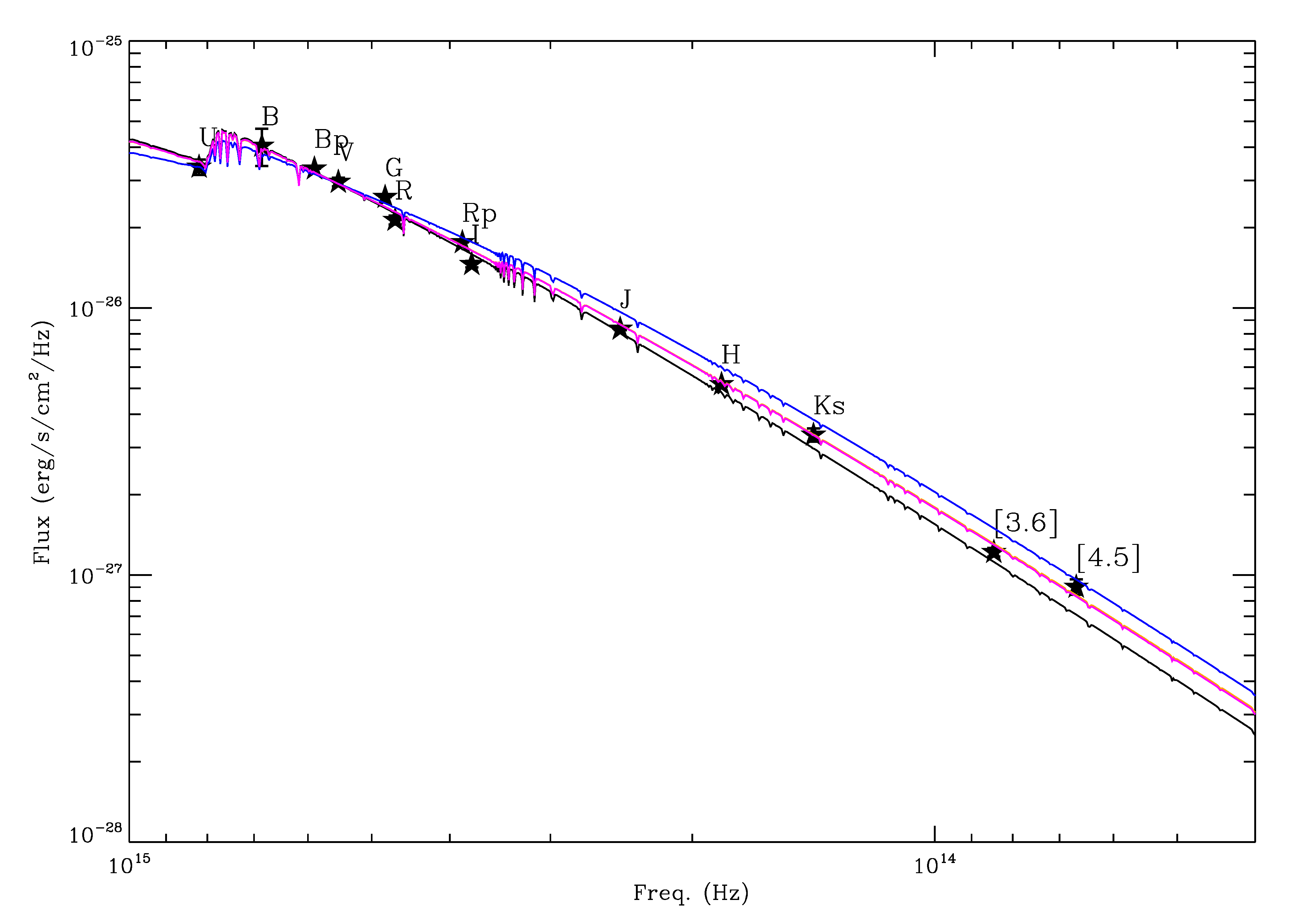}
    \caption{Comparison of model SEDs (continuous lines) with extinction corrected photometric data (stars with labels, error bars are small and difficult to see on this scale). Models are normalized to the $V$-band and parameters are adjusted to fit the IR excess as noted in 4.3. Color coding as follows: black -- composite stellar spectrum; magenta -- flat Be disk and accretion disk model (both models are essentially identical on this scale); blue -- for comparison this illustrates a flat Be disk model if it were circumstellar to the B-type star and truncated at its Roche radius. }
    \label{fig:sed}
\end{figure}

The compactness of the disk, as implied by the VLT-FLAMES data, is consistent with the very faint IR excess observed in the SED, with an appropriate choice of uncertain disk parameters, see Fig.~\ref{fig:sed}. For example a flat Be-star disk model \citep{carciofi2006} reproduces the observed IR excess for an inner disk temperature of $\sim$30\,000$\pm10\,000$\,K and inner/outer disk radii of 0.7/7.0\,$R_\odot$. A simple $\alpha$-disk accretion model \citep{shakura,beall} requires an accretion rate of $10^{-7}\,M_\odot yr^{-1}$ in order to reproduce a similar IR excess, with a BH mass of 25\,$M_\odot$. Truncating the outer radius at 7\,$R_\odot$ reduces the IR excess significantly, hence the high accretion rate needed to match the observations. The latter would imply an X-ray luminosity several orders of magnitude in excess of the observed limit from XMM-Newton of $L_x<5.3\,10^{33}$\,erg\,s$^{-1}$, and is also far above the expected accretion rate of such a system.
For example, a B-type star with the primary's parameters is expected to have a mass-loss rate of $\sim 10^{-10} M_\odot$yr$^{-1}$ \citep{krticka} giving an accretion rate onto the BH roughly an order of magnitude smaller than this. Decreasing the accretion rate in the model to this magnitude would require increasing the BH mass by some orders of magnitude to recover the IR excess. 
However, at such a low accretion rate the accretion may proceed via an advection dominated accretion flow (ADAF) as is the case for BH X-ray binaries at low accretion rates \citep{esin,narayan}. 
While producing ADAF models is beyond the scope of this paper, and in any case we only have upper limits on the X-ray flux, such models might succeed in matching the small IR excess, without exceeding the X-ray limiting flux. 
In summary, while the weak IR excess can be modeled with a disk size within the suggested limits, the nature of that disk is still obscure.

Implicit in the above discussion is the assumption that the disk is in equilibrium. From the observations this is clearly not the case, though the strong Balmer emission in the VLT-FLAMES data is present for $\sim$60 days, or roughly 30 periods of the inner binary.  However, considering circumbinary disks, binaries tend to clear the inner disk to a distance of $\sim2a$ on dynamical timescales \citep{artymowicz,rudak1981}, that is, within a few tens of periods. It is possible, therefore, that the disk is episodic and the emission observed at later times, in the SALT-HRS data, is a remnant circumbinary structure.  We assume that the tertiary is essentially a spectator in this scenario.

\subsection{Constraints from the tertiary}

Since our knowledge of the tertiary star's properties are only approximate we can only make some broad inferences. The $V$-band flux ratio yields an absolute magnitude of M$_{\rm V}=-2.5\,m$ for the tertiary and, assuming it lies on an isochrone with the narrow-lined B-type star, implies a mass of approximately 6--7\,$M_\odot$. While the longer period is very poorly sampled, the VLT-FLAMES data imply that a period of less than $\sim$120\,d is inconsistent with the data. If we assume that spin and orbital angular momentum of the tertiary are aligned then its high \vsini\ argues for a significant tilt of its orbital plane with respect to the inner binary. Assuming an equatorial rotational velocity close to the escape velocity of the star provides lower limit of sin$i$>0.5.

\begin{figure*}
    \centering
    \includegraphics[width=\hsize]{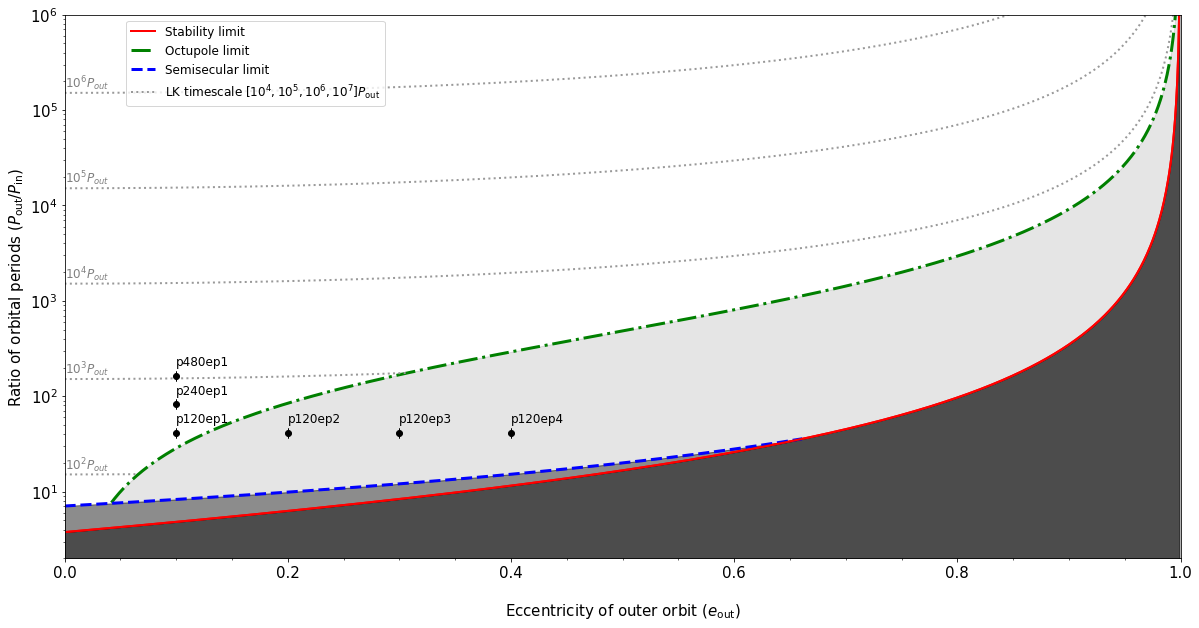}
    \caption{Visualization of stability criteria for the triple systems \citep[][where further details can be found]{toonen2020}, the shaded area below the red line indicating unstable configurations. The locations of test models are illustrated, assuming $m_1$=25, $m_2$=8.6, $m_3$=6\,$M_\odot$, and $P_{in}=2.92$\,d. The range of outer periods and eccentricities are indicated by the labels, for example the model p120ep1 assumes $P_{out}$=120\,d and an eccentricity of 0.1 for the outer system. All test models assume prograde orbits of \ang{87} inclination of the orbital planes.}
    \label{fig:triple}
\end{figure*}

The stability of such a triple star system can be investigated following \citet{toonen2020} and is illustrated in Fig.~\ref{fig:triple}\footnote{Fig.~\ref{fig:triple} was constructed with the python notebook at https://bndr.it/wr64f}. For these tests we adopted an inclination angle between the inner and outer binaries of \ang{87} in order to show the limit of secular instability. Changing the inclination such that it differs from \ang{90}, the angle that maximizes the secular limit extent, by only a few degrees more shrinks the secular limit to negligible size. We see that stability increases with increasing outer period and decreasing eccentricity and, from the above, that the limit of secular instability is only significant for inclinations very close to \ang{90}. 

It is difficult to be more specific except to note that the velocity dispersion of the radial velocity measurements of the proposed tertiary are larger than those of the emission line velocities, and the possible range of inner binary systemic velocities (see Fig.~\ref{fig:solutions} and Tables \ref{table:broadrv} and \ref{table:emrv}). Naively, one might interpret this as support for the BH hypothesis, though there are significant uncertainties for the tertiary results.

Further radial velocity monitoring would help clarify the nature of the system. For example, the primary has such narrow metal lines that radial velocity measurement is straightforward. Hence spectroscopic monitoring with a cadence covering the inner period, and an appropriate overarching cadence to sample potential outer orbital periods would enable two complementary methods. On the one hand, the radial velocities of the tertiary may be retrievable with sufficient accuracy to determine the mass ratio of the inner binary to the tertiary directly, thus enabling a determination of m$_2$. On the other hand, should that not be feasible, the systemic radial velocity data of the inner binary alone may permit determination of m$_2$ provided one can determine the mass function ($f$) for the system consisting of the inner binary and the tertiary, and hence solve the resulting quadratic equation to give
\[  {\rm m}_1+{\rm m}_2=\sqrt{{\rm m}^3_3{\rm sin}^3i/f} - {\rm m}_3\] subject of course to eccentricity and orbital plane considerations.
An interesting corollary is that triple, or higher order, systems may provide an unambiguous method for determining the masses of suspected dark or subluminous members, depending on their periods.

\section{Conclusions and caveats}

We determine NGC\,2004\#115 to be a triple system comprising an inner short-period (2.92 d) system with a very low \vsini\ 8.6\,$M_\odot$ B-type primary star and, assuming synchronous rotation, a BH of mass $>25\,M_\odot$. The tertiary is determined to be a broad-lined 6--7\,$M_\odot$ B-type star with an outer period in excess of 120 days.
No sign of a secondary spectrum is observed in the inner binary. 
The velocity structure of the early time doublepeaked H$\alpha$ emission line is consistent with an accretion disk inside the Roche lobe of the secondary, and its centroid velocity is remarkably constant.  
It is therefore useful to consider the plausibility of forming an inner binary consisting of an 8.6\,$M_\odot$ main sequence stars and a 25\,$M_\odot$ BH.

While the current system is relatively stable (from Fig.~\ref{fig:triple}) we have no knowledge of its prior state.
Nevertheless, \citet{toonen2021} recently presented an overview of the evolution of triple systems and showed that, ignoring the small percentage of disrupted systems, roughly three quarters of triples follow a path in which the primary star of the inner binary initiates mass transfer. The remaining evolve with no interaction except for a small percentage that follow paths in which either the secondary or tertiary initiates mass transfer, each comprising of roughly 1\% of the total. We thus consider the evolution of the inner binary in isolation.

Assuming the BH mass equals the helium core mass at collapse implies a progenitor star of at least 50\,$_\odot$ \citep{Woosley_2019}.
This makes an initially short orbital period ($\lesssim$20\,d) rather unlikely, since most systems with such extreme initial mass ratios (<0.2) cannot undergo stable mass transfer, and would therefore quickly merge \citep[cf Fig,\,B1 in][]{langer}. Binaries that are initially wider may undergo stable mass transfer for extreme mass ratios if the mass transfer is inefficient; however, the post-mass transfer orbital periods would generally be larger than 10\,d.
Furthermore, the normal surface abundances of the inner B-type star argue against significant mass transfer from the primary.

Common envelope (CE) evolution may provide a more viable explanation. Stars of 50\,$M_\odot$ or more are expected to go through an LBV phase \citep{smith2011}, and lose their hydrogen-rich envelope even as single stars. Therefore it is plausible that an 8.6\,$M_\odot$ companion in an orbit at several 100\,R$_\odot$ separation could survive CE evolution, but be dragged into a short period orbit \citep{kruckow2016}. Alternatively, the inner binary could result from a tight extreme mass ratio binary in which only the massive component evolves chemically homogeneously. According to \citet{Marchant_2017}, while this scenario requires slightly sub-LMC metallicity, it may become possible by enhanced mixing in the closest binaries \citep{hastings2020,koenigsberger2021}, which might also explain the shortest period Wolf-Rayet binaries in the LMC \citep{Shenar_2020}. 

An important caveat is that the inclination hinges on the assumption that the inner binary is in synchronous rotation, the upper limit on sin$i$ of 0.15 (\ang{9}) following from the period of the system and the radius of the primary B-type star.  The period is well determined, in spite of the incomplete phase coverage of the VLT-FLAMES data, and leaves little room for manoeuvre. The radius is determined by the assumption that the star must have LMC composition and the system resides in the LMC. The latter condition is fully consistent with its negligible parallax, proper motion, systemic radial velocity and presence of LMC-like interstellar lines. A substantially smaller radius would imply significant helium enhancement \citep{lennonlb1} accompanied by a very strong C/N depletion/enhancement. The latter is not supported by the data.
Furthermore, as the cirularization time of the inner binary, $\sim$16\,Myr, is much larger than its synchronization time, 0.07\,Myr, this strongly supports the assumption of synchronicity. Indeed an 8.6\,$M_\odot$ star reaches a radius of 5.6\,$R\odot$ after about 20\,Myr \citep[H-exhaustion occurs after 32\,Myr,][]{brott}. This picture would be consistent with the BH-progenitor initiating interaction after about 4\,Myr, the lifetime of a massive star.

Nevertheless, an inclination angle of \ang{9} or lower has a probability of about 1\%, hence making the ad hoc assumption that the inner binary is asynchronous, there are two options: If the star is rotating more quickly then $i$ must be smaller than \ang{9}, the BH mass more massive, and we are entering into an even less probable scenario. If the star is rotating more slowly, $i$ is larger and the BH mass is smaller. Indeed if $i$ is large enough we enter the range where the companion might even be a more mundane intermediate mass star. In this case, the primary would have a very low equatorial velocity and hence would need to lose most of its angular momentum. The most feasible way of achieving this is via Roche lobe overflow, leading to the formation of a stripped star. However, as already demonstrated, this solution is not viable. Furthermore, if the secondary would indeed be a main sequence star of similar mass to the component identified as the tertiary, its Balmer line radial velocity variations would be substantially larger than those already identified for the tertiary.

In conclusion, while our investigation into the nature of this enigmatic system leads to the surprising result that it may harbor a $\sim$25\,$M_\odot$ black hole, further spectroscopic monitoring is required on cadences that will provide stronger constraints on the both inner binary and the proposed tertiary. Such data might also be amenable to spectral disentangling techniques.

During the refereeing stage of this paper we became aware that \citet{saracino} reported the discovery of a B+BH binary in the LMC cluster NGC1850, consisting of a 5\,$M_\odot$ B-type star plus an 11\,$M_\odot$ BH. However \citet{elbadry1850} have argued that the primary star is a stripped star with a mass of only $\sim$1\,$M_\odot$, possibly with a main sequence companion.  Applying our methodology (cf Fig.~\ref{fig:meq8})
to the B+BH system 
we estimate a stellar radius of 6.6\,$R_\odot$
and a stellar luminosity of $\log L/L_\odot$=3.2. The mass function yields a Roche radius of approximately 10\,$R_\odot$, giving a fill factor of only 0.66. This system cannot reproduce the observed light curve as it would generate maximum (i.e., sin$i$=1) ellipsoidal light variations with an amplitude of only $\sim$4--5\%, significantly smaller than the observed light curve constraint of $\sim$10--12\%, in broad agreement with \citet{elbadry1850}. 
Reducing the mass of the primary
%, still with the constraint that it contributes 100\% of the light, 
we find that its Roche radius matches the stellar radius for a mass of $\sim$1.8\,$M_\odot$, the light curve then implying i$\sim$\ang{47} with a BH mass of $\sim$5.8\,$M_\odot$.
However if a secondary contributes significantly to the light then the radius and luminosity of the primary would be smaller, and the intrinsic ellipsoidal light variations would be larger, thus permitting an even lower mass primary, and secondary, as a possible solution. In this case the mass and luminosity of the primary would imply that it is a stripped star, as also found by \cite{elbadry1850}.
As described in the present paper, a good quality high resolution spectrum may enable a detailed abundance analysis and a direct determination of the flux contribution and radius of the narrow-lined B-type component.

\begin{acknowledgements}

DJL acknowledges support from the Spanish Government Ministerio de Ciencia, Innovaci\'on y Universidades through grants PGC-2018-091 3741-B-C22 and from the Canarian Agency for Research, Innovation and Information Society (ACIISI), of the Canary Islands Government, and the European Regional Development Fund (ERDF), under grant with reference ProID2017010115. DJL thanks Ian Howarth for providing high resolution ATLAS spectra for the LMC from \citet{howarth2011} that were instrumental in looking for cool companions in the inner binary.
The SALT data were acquired under the auspices of Director's Discretionary Time proposal 2020-1-DDT-007 (PI: I. Monageng).
We thank the referee, Tomer Shenar, for a careful reading of the manuscript that led to a significant improvement of the paper.
This work has made use of data from the European Space Agency (ESA) mission
{\it Gaia} (\url{https://www.cosmos.esa.int/gaia}), processed by the {\it Gaia}
Data Processing and Analysis Consortium (DPAC,
\url{https://www.cosmos.esa.int/web/gaia/dpac/consortium}). Funding for the DPAC
has been provided by national institutions, in particular the institutions
participating in the {\it Gaia} Multilateral Agreement.
This research  made use of the SIMBAD and VIZIER databases (operated at CDS, Strasbourg, France) and the table manipulation software TOPCAT \citep{topcat}.

\end{acknowledgements}

\bibliographystyle{aa}
\bibliography{references.bib}

\begin{appendix}

\section{Radial velocity measurements}
\label{appendix:rv}

\subsection{VLT-FLAMES data}

\begin{figure}
    \centering
    \includegraphics[width=\hsize]{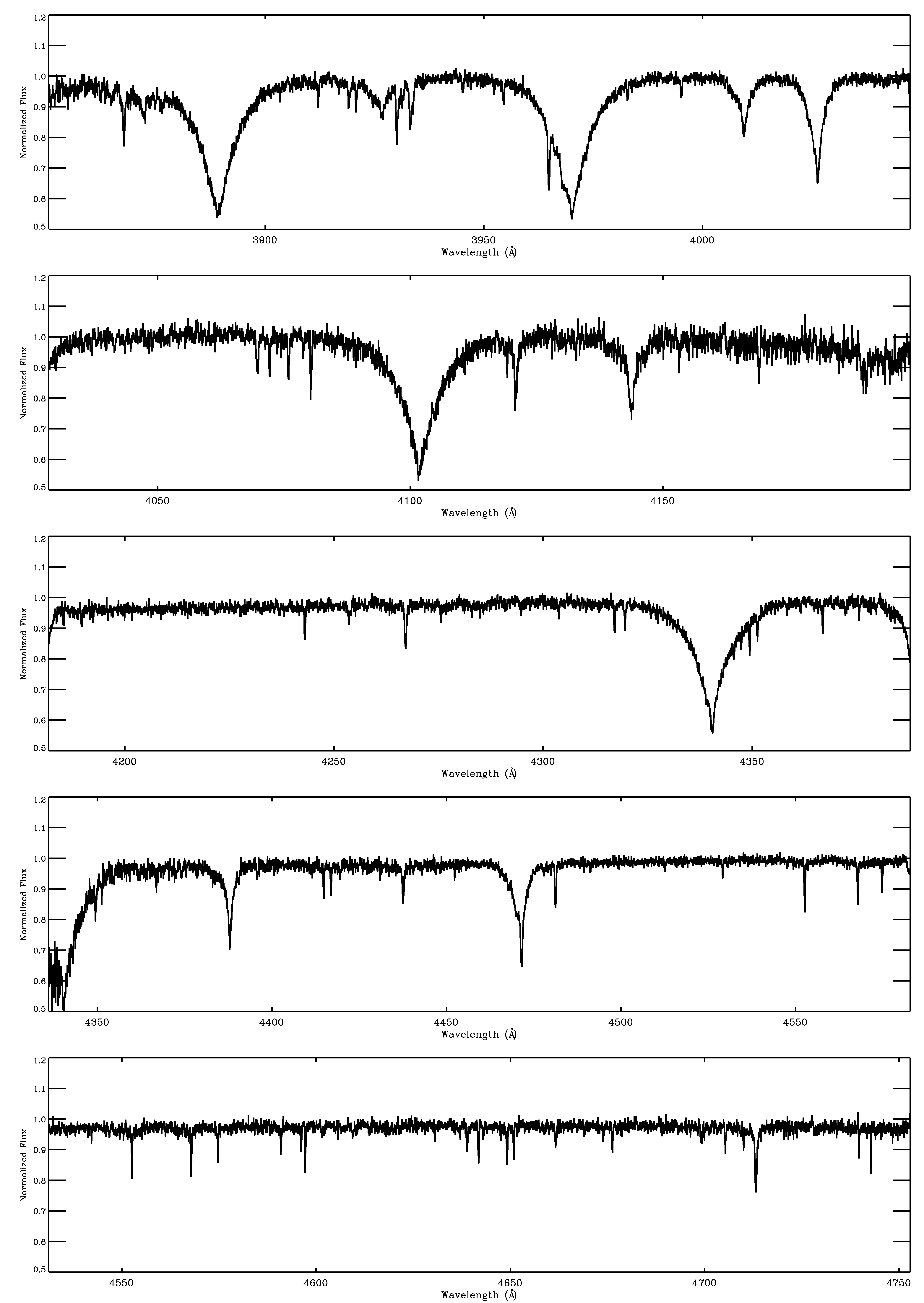}
    \caption{VLT-FLAMES merged data for the spectral regions, top to bottom, HR02, HR03, HR04, HR05 and HR06. Approximate normalization of the spectra was accomplished with a low order polynomial, for line profile fitting a custom normalized was performed.}
    \label{spectra:primary}
\end{figure}

Radial velocities of the narrow lined B-type star were measured by cross-correlating each spectrum against an LMC model template of similar spectral type (an exact match is not required), convolved with the instrumental resolution \citep[from][]{evans2006} and appropriate \vsini. 
Where possible we used the metal lines to measure the radial velocities, selecting segments of the spectrum in each of the observed FLAMES settings that contain sufficient strong metal lines to produce a strong correlation peak, avoiding the hydrogen and helium lines that may be broad, asymmetric or have a contribution from the disk. Note that \ion{He}{i} lines were used for the HR02 observations due to the lack of suitable strong metal lines therein.
The cross-correlation segment regions are listed in Table \ref{table:A.1}, together with a summary of the metal species included in those segments. 
As noted in Sect.~2, each of the six regions was observed in two observing blocks (OB), OB1 and OB2, with three separate exposures in each OB, exposures 1--3, and 4--6. 
Radial velocities were measured for each individual exposure except for the HR03 observations as exposures 5 and 6 of OB2 had very low signal levels and were not useful. The blue spectrum, illustrated in Fig.~\ref{spectra:primary}, is a typical sharp lined B2 spectrum, the B2e classification coming from the peculiar H$\alpha$ emission line shown in Fig.~\ref{fig:halpha}.

Radial velocities were measured by approximating the cross-correlation peak with a Gaussian function, a good approximation for the narrow metal lines, and determining its position using a least squared fit to its width, position and base level. The 1-$\sigma$ uncertainty in the position is adopted as the uncertainty in the radial velocity. 
The extreme narrowness of the lines leads to typical uncertainties of less than 1\,\kms\ for each spectrum, though the systematic velocity precision from this instrument is estimated as $\sim$2\,\kms\ between grating settings \citep{evans2008}.  
The results are given in Table \ref{table:A.2} and, depending on epoch, the short period enables one to discern small velocity shifts within the separate exposures of certain OBs.

As a check of these values we also determined the maximum of the peak position of the cross-correlation function as a measure of the radial  velocity. 
Both methods were found to give very similar results, with agreement within the errors, except for the HR02 data where small shifts of up to 5\,\kms\ are observed between these two methods that we attribute to the use of \ion{He}{I} lines that are slightly asymmetric and skew the Gaussian fit away from the peak position.

\begin{table}[h]
\caption{Wavelength ranges covered by FSMS instrument settings and spectral segments used for cross-correlation analysis and the strongest absorption lines in each region (for the six different FLAMES settings). Note that the HR05 and HR06 observations overlap, so the same wavelength range and lines were used for both regions.}             % title of Table
\label{table:A.1}      
\centering                        
\begin{tabular}{cccl}       
\hline\hline                
FLAMES & Wavelength   & Segment & Primary species \\   
setting & range (\AA) & range (\AA) & \\
\hline                 
  HR02  & 3854--4051 & 4003--4041 & \ion{He}{i} 4009,4026 \\ 
  HR03  & 4032--4203 & 4065--4095 & \ion{O}{ii} 4070,4072,\\
  & & & 4076,4079\\
  HR04  & 4187--4394 & 4260--4280 & \ion{C}{ii} 4267, \\
  & & & \ion{O}{ii} 4274,4276\\
  HR05  & 4340--4587 & 4550--4585 & \ion{Si}{iii} 4553,4568,\\
  & & & 4575 \\
  HR06  & 4537--4760 & 4550--4585 & \ion{Si}{iii} 4553,4568,\\
  & & & 4575 \\
  HR14A & 6308--6701 & 6580--6600 & \ion{C}{ii} 6578,6583 \\
\hline                                   %inserts single line
\end{tabular}
\end{table}

\begin{table}
\caption{Results of the radial velocity analysis of the FSMS and SALT data.}   
\label{table:A.2}      
\centering                        
\begin{tabular}{r c l c}       
\hline\hline 
Instr. & Exp. & Rad. vel.& Date \\
setting & no. &  \kms  &    MJD \\
\hline
HR02 OB1 &  1 & $272.56   \pm 0.44  $& 52982.20114 \\
     &  2 & $ 282.45  \pm 0.44  $& 52982.22809 \\
     &  3 & $ 285.00  \pm 0.45  $& 52982.25504 \\
OB2     &  4 & $ 294.04  \pm 0.44  $& 52988.21908 \\
     &  5 & $292.32   \pm 0.46  $& 52988.25059 \\
     &  6 & $296.05   \pm 0.45  $& 52988.27754 \\
\\
HR03 OB1 &  1 & $ 262.97  \pm  0.81 $& 53005.13673 \\
     &  2 & $ 265.51  \pm  0.75 $& 53005.16375 \\
     &  3 & $ 266.86  \pm  0.72 $& 53005.19070 \\
OB2     &  4 & $ 270.96  \pm  0.80 $& 53005.22133 \\
     &  5 &  -- 		  & 53005.24835 \\
     &  6 &  -- 		  & 53005.27531 \\
\\
HR04 OB1 &  1 & $ 369.75  \pm  0.82 $& 53006.13530 \\
     &  2  &$  376.80  \pm  0.71$& 53006.16225 \\
     &  3  &$  377.35  \pm  0.76$& 53006.18921 \\
OB2     &  4  &$  268.31  \pm  0.90$& 53008.04798 \\
     &  5 & $ 264.53  \pm  0.90 $& 53008.07501 \\
     &  6 & $ 267.57  \pm  0.79 $& 53008.10197 \\
\\
HR05 OB1 &  1 & $ 269.73  \pm  0.37 $& 53008.16305 \\
     &  2  &$  271.66  \pm  0.32$& 53008.19000 \\
     &  3  &$  275.40  \pm  0.48$& 53008.21695 \\
OB2     &  4  &$  377.29  \pm  0.40$& 53009.16182 \\
     &  5 & $ 381.95  \pm  0.35 $& 53009.18877 \\
     &  6 & $ 382.36  \pm  0.42 $& 53009.21573 \\
\\
HR06 OB1 &  1 & $ 382.10  \pm  0.37 $& 53012.09557 \\
     &  2  &$  383.44  \pm  0.35$& 53012.12259 \\
     &  3  &$  383.12  \pm  0.46$& 53012.14954 \\
OB2     &  4  &$  385.96  \pm  0.43$& 53012.18344 \\
     &  5 & $ 385.79  \pm  0.46 $& 53012.21045 \\
     &  6 & $ 387.76  \pm  0.59 $& 53012.23274 \\
\\
HR14A OB1&  1 & $ 246.56  \pm  0.62 $& 52955.26626 \\
     &  2 &$  244.62  \pm  0.72$& 52955.29327 \\
     &  3 &$  242.14  \pm  0.65$& 52955.32022 \\
OB2     &  4 &$  371.88  \pm  0.55$& 52989.25927 \\
     &  5 & $ 373.07  \pm  0.64 $& 52989.28623 \\
     &  6 & $ 371.48  \pm  0.70 $& 52989.31318 \\
     \\
SALT/HRS  &  1 & $ 396.6 \pm 0.7 $ & 59110.08144 \\
  &  2 & $297.5 \pm 0.9 $& 59112.06897 \\
  &  3 & $312.5 \pm 0.7$ & 59115.09170 \\
  &  4 & $389.6 \pm 0.6$ & 59116.07530 \\ 
\hline                                   %inserts single line
\end{tabular}
\end{table}

\subsection{SALT-HRS data}

The HRS instrument on SALT is a dual blue-arm/red-arm echelle spectrograph that was used to cover the spectral regions 3730--5570 and 5420--8790\AA\ at a spectral resolution of R$\sim$\num{14000}. The S/N of the observations for the blue arm data ranged from 20--40 per pixel, though it dropped very rapidly below 4500\AA\ such that the lines in this region were not useful. The S/N for the red arm data was more uniform and ranged from 20--30 per pixel, although the night sky lines were very strong. This latter issue was a concern for the H$\alpha$ region and it was decided not to subtract the sky lines from the data to avoid unknowingly distorting the profile. Due to the low S/N of the SALT data radial velocities were determined using a Gaussian fitting approach to lines belonging to the \ion{Si}{III} triplet at 4553/4568/4575\,\AA\ and the \ion{C}{ii} doublet at 6578/6383 \AA. The low s/n in the blue region precluded the use of metal lines below 4500\AA. Results are also listed in Table \ref{table:A.2}.

\section{Atmospheric analysis of the B-type spectrum}
\label{appendix:atmosphere}

Typically, \teff and \logg are determined by satisfying one or more of the ionization balances \ion{Si}{ii}/\ion{Si}{iii},  \ion{Si}{iii}/\ion{Si}{iv} or \ion{He}{i}/\ion{He}{ii}, while simultaneously fitting Balmer lines such as the H$\gamma$ and H$\delta$ profiles. In the case of NGC\,2004\#115, \ion{Si}{iii} is the only ionization species that is present with significant strength, both \ion{Si}{ii} and \ion{Si}{iv} are marginally present, while \ion{He}{ii} is absent. Nevertheless the weakness of the \ion{Si}{ii} and \ion{Si}{iv} lines provide strong constraints on the effective temperature, but the Balmer lines are compromised due to the presence of another component preventing their use as the main \logg\ constraint.

Therefore, for a given value of \logg\ we determined the effective temperature and element abundances by comparing observed equivalent widths of the metal lines, Table~\ref{table:ews}, with precomputed grids.
Table~\ref{table:ews} includes the \ion{C}{ii} doublet at 6580\AA that was not included in previous FSMS analyses, but has been found by \citet{barnett1988} to provide reliable carbon abundance estimates. Excluding this doublet would change the estimates by less than 0.05 dex. 
The veiling effect of second light is approximated by a stellar continuum with \teff=20\,000\,K although changing this temperature by $\pm5\,000$\,K has little impact on the final results. This continuum is scaled to the required $V$-band flux ratio and used to rescale the predicted equivalent widths to diluted values for comparison with observation. The overarching procedure then consists of running a sequence of calculations at a given \logg\ to find the \teff\ and flux ratio that reproduces the normal LMC silicon abundance, the results of which are presented in Fig.~\ref{fig:fraction}. Further details on the method can be found in \citet{lennonlb1}.

The microturbulent velocity (\vt) deserves special mention. It was estimated from the relative strength of the components of the \ion{Si}{iii} triplet near 4560\AA, a positive abundance gradient with equivalent width indicating too low a value, a negative gradient, too high a value. With no veiling, a negative slope is found for all values of the \vt, including \vt=0\,\kms. This anomaly has been identified previously by \citet{hun07} for a small number of FSMS targets, with one explanation being veiling by an unidentified companion. For larger veilings ($\sim$50\%), the observed equivalent widths are consistent with \vt$\simeq$0 within their estimated errors. Hence we conclude that the microturbulent velocity is small with our best estimate being 0\,\kms, though with low significance as values up to 2--3\,\kms\ are difficult to exclude. The main impact of an increased \vt\ is to increase the veiling (i.e., decrease the contribution of the B-type star to the total flux), for example with \vt=2\,\kms\ the B-type star would be fainter by about 6\% of the total flux.

The projected rotational velocity, \vsini, was estimated from the \ion{Si}{iii} 4560\AA\ triplet which was observed with both the HR05 and HR06 settings, providing six independent estimates. Theoretical spectra for the best fitting atmospheric parameters listed in Table \ref{table:t_abund} were convolved with the instrumental \citep[assumed to have a Gaussian profile and with FWHMs taken from][]{evans2006} and rotational profiles, and then re-binned to match the observations. In general, the observed profiles are best fitted with \vsini$\sim$10\,\kms, with the estimates being little affected by changes in the adopted effective temperature or gravity. Increasing the microturbulent velocity to 2\,\kms\ would have reduced the estimates by typically 2\,\kms. Similarly changing the FWHM of the instrumental profile by 10\% would typically change the \vsini\ estimates by 1\,\kms. In summary our best estimate of the projected rotational velocity is 10\,\kms. Allowing for possible systematic uncertainties leads to a robust upper limit, \vsini$\leq$15\,\kms.

\begin{table}
\caption{Metal line identifications and equivalent widths, with errors. Lines that are significant blends with other components are indicated in the Comment field. }
\centering
\begin{tabular}{lcrl}\hline\hline 
Ion  & wavelength & ew & Comment \\ 
 & (\AA) & (m\AA) & \\ \hline
\ion{C}{ii}   &  3919 & $23\pm2.2$ & \\
\ion{C}{ii}   &  3921 & $32\pm2.5$ & \\
\ion{C}{ii}   &  4267 & $73\pm3.0$ & doublet \\
\ion{C}{ii}   &  6578 & $58\pm1.9$ & \\
\ion{C}{ii}   &  6582 & $42\pm1.8$ & \\
\ion{N}{ii}   & 3995  & $23\pm2.0$ & \\
\ion{N}{ii}   & 4447  & $10\pm2.0$ & \\
%\ion{N}{ii}   & 4601  & 9   & \\
\ion{N}{ii}   & 4630  & $13\pm1.5$ & \\
\ion{N}{ii}   & 4643  & $11\pm1.9$ & \\
\ion{O}{ii}   & 3845  & $18\pm2.3$ & \\
\ion{O}{ii}   & 3954  & $25\pm2.5$ & \\
\ion{O}{ii}   & 4070  & $50\pm7.0$ & \ion{O}{ii} blend \\
\ion{O}{ii}   & 4072  & $27\pm2.8$ & \\
\ion{O}{ii}   & 4076  & $35\pm2.8$ & \\
\ion{O}{ii}   & 4079  & $11\pm2.1$ & \\
\ion{O}{ii}   & 4317  & $26\pm1.6$ & \ion{O}{ii} blend \\
\ion{O}{ii}   & 4320  & $26\pm1.7$ & \ion{O}{ii} blend \\
\ion{O}{ii}   & 4367  & $24\pm2.0$ & \ion{O}{ii} blend \\
\ion{O}{ii}   & 4415  & $33\pm2.0$ & \\
\ion{O}{ii}   & 4417  & $31\pm2.0$ & \\
\ion{O}{ii}   & 4452  & $10\pm1.5$ & \\
\ion{O}{ii}   & 4591  & $27\pm1.8$ &  \\
\ion{O}{ii}   & 4596  & $23\pm1.9$ & \\
\ion{O}{ii}   & 4639  & $29\pm2.8$ & \\
\ion{O}{ii}   & 4642  & $38\pm1.8$ & \\
\ion{O}{ii}   & 4649  & $47\pm1.9$ & \\
\ion{O}{ii}   & 4651  & $24\pm1.6$ & \\
\ion{O}{ii}   & 4661  & $27\pm2.1$ & \\
\ion{O}{ii}   & 4676  & $23\pm1.7$ & \\
\ion{O}{ii}   & 4705  & $19\pm2.7$ & \\
\ion{O}{ii}   & 4710  & $13\pm1.7$ & \\
\ion{Si}{ii}  &  4128 & $8\pm2.9$ & \\
\ion{Si}{ii}  &  4130 & $6\pm2.9$ & \\
\ion{Si}{iii} &  4553 & $57\pm2.5$ & \\
\ion{Si}{iii} &  4568 & $47\pm1.8$ & \\
\ion{Si}{iii} &  4575 & $34\pm1.7$ & \\
\ion{Si}{iv}  &  4089 & $6\pm2.5$ & \ion{O}{ii} blend\\
\ion{Si}{iv}  &  4116 & $7\pm2.5$ & \\
\ion{Mg}{ii}  & 4481  & $65\pm2.5$ & doublet \\
\hline
\end{tabular}
\label{table:ews}
\end{table}
% Checked by PLD 24/8/21

\section{Limits on a secondary spectrum}
\label{appendix:secondary}

\begin{figure*}
    \centering
    \includegraphics[width=0.9\hsize]{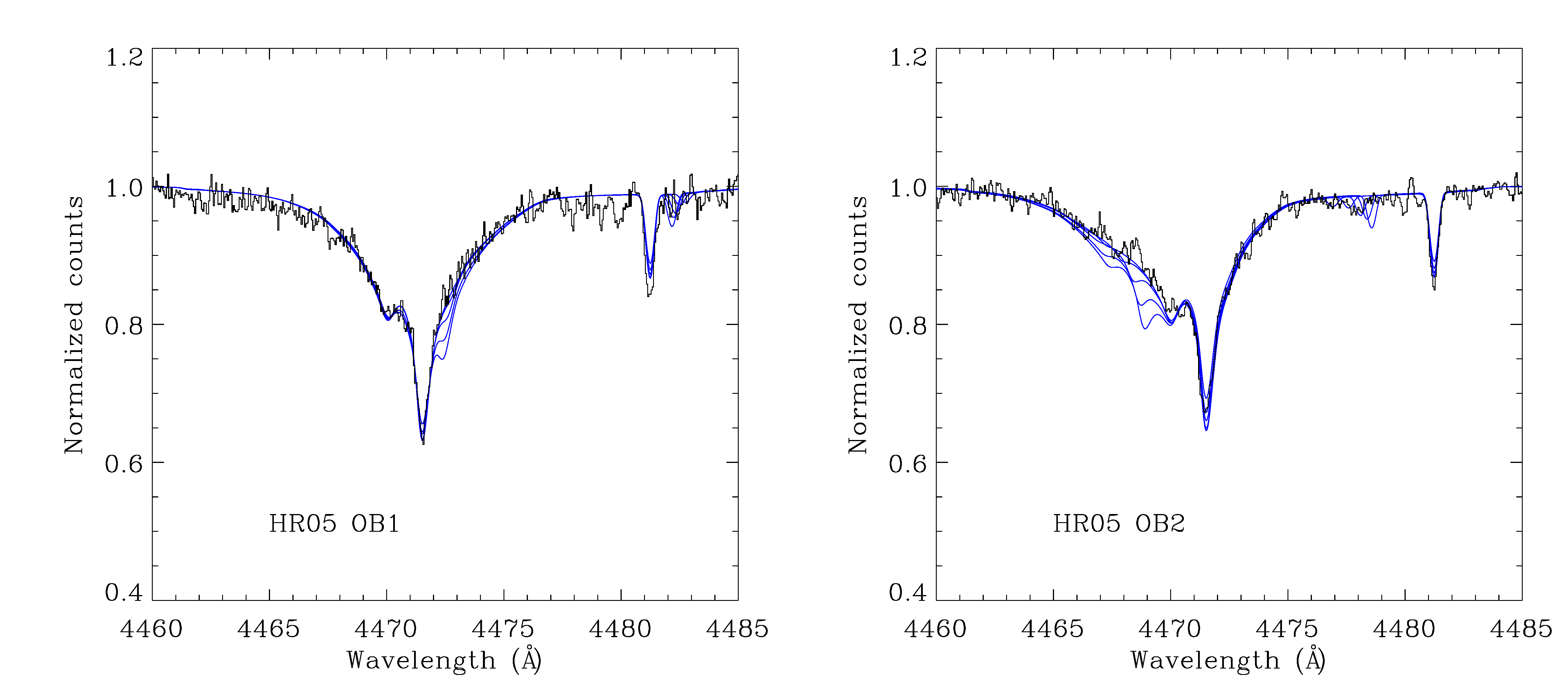}
    \caption{Comparison of observed data for HR05 OB1 and OB2 (black lines) with simulations of the sum of primary, secondary and tertiary spectra (blue lines) weighted according to their absolute visual magnitude, $M_V$, from Table\,\ref{tab:C.1}. The main features are the \ion{He}{i} 4471\,\AA\ and \ion{Mg}{II} 4481\,\AA\ lines. The former shows no evidence of the secondary implying that the spectral type must be B3\,V or later. The constraint derived from the \ion{Mg}{II} is discussed in the text.}
    \label{fig:app_c}
\end{figure*}

Here we examine whether the secondary, if it is a main sequence star, would be visible in our data. We make the assumption that this secondary is also narrow lined, by virtue of the expected short synchronization times. Since a narrow-lined secondary is not observed in our data this implies that the radius of the secondary is smaller than the primary, and its \vsini\ should be smaller.
Hence, its spectrum would not be resolved with the present data and we therefore assume a line width determined by instrumental resolution of FWHM$\sim$15\,\kms. 
A broad lined, or quickly rotating, star would clearly be much more difficult to detect. 
We adopt notional effective temperatures and luminosities for a range of near main-sequence stars, Table~\ref{tab:C.1}, and assign them spectra from the TLUSTY NLTE grid discussed in Appendix \ref{appendix:atmosphere}.
Expected masses are also listed in this table for two different age assumptions, where we have used the zero rotation isochrones of \citet{mowlavi}, as the tracks of \citet{brott} do not extend to masses below 5\,$M_\odot$.  Note that evolution has little impact for the fainter stars, the same being true for rotation, and indeed the exact choice of stellar evolution models.

As more massive and hotter stars would be easily identified due to their contribution to the hydrogen Balmer lines and \ion{He}{i} lines (Fig.~\ref{fig:app_c}), we focus on features that provide a strong contrast for the cooler stars.
The \ion{Mg}{ii} doublet at 4481\,\AA\ is well suited to this purpose. It is a strong feature that has a maximum strength on the main sequence at roughly A0 spectral type, hence strengthening as one moves to cooler, less massive stars.
There are other candidates, such as the \ion{Si}{ii} doublet at 6347,6371\,\AA\ and the \ion{Ca}{ii} doublet at 3934\,\AA\ (the 3968\,\AA\ component is blended with H$\epsilon$). However, \ion{Mg}{ii} is preferred as the  signal-to-noise in this region is superior, it lies in a relatively clean part of the spectrum, and moreover the relative radial velocities of OB1 and OB2 are significantly different from zero; approximately -28 and +80 \kms\ respectively relative to the systemic velocity (values are adopted from the orbital solution).  The negative shift of the primary is particularly useful, as the proposed tertiary is also blue shifted at these epochs, hence the secondary should be red-shifted at that time. 

Synthetic secondary spectra were computed assuming a radial velocity appropriate for the assumed mass, weighted by their respective magnitudes, corrected to distance of the LMC, and then diluted with a featureless continuum representing the tertiary plus primary such that the observed dereddened magnitude of the system is conserved.
The equivalent width of the resulting \ion{Mg}{ii} line of the secondary was then measured and, since no such component has been identified, it was compared with the detection limit (limiting equivalent width) of the data.  
Since the secondary component has different positions in OB1 and OB2 we shifted both OBs to the rest frame of the potential secondary, then merged the spectra and measured the limiting equivalent width in the resulting spectrum. The results are summarized in Table C.1 and can be compared with the 1-$\sigma$ detection limit in this region of the spectrum that we estimate as 2.6\,m\AA\ (based on 1000 equivalent width measurements at random continuum positions). We can therefore exclude, at a 2-$\sigma$ confidence level, the presence of a secondary main sequence star with an absolute visual magnitude above $\sim$+0.6.  If the star would be a zero age main sequence star its mass would be $\sim$3.3\,$M_\odot$, a more evolved star would be correspondingly less massive.

Allowing higher values of \vsini\ for the secondary obviously makes it harder to detect. Values of 100 and 200\,\kms\ imply 1-$\sigma$ detection limits of 6.6 and 8.8\,m\AA\ respectively for this line. On the other hand the brighter stars, B3\,V and above, are more easily detected using their stronger \ion{H}{i} and \ion{He}{i} lines for these moderate \vsini\ values. 

\begin{table}
    \caption{Adopted parameters for notional main sequence secondary stars, we assume \logg=4.25 for all models. We also list the resulting effective equivalent width for the \ion{Mg}{ii} 4481\,\AA\ line in the combined spectrum. Estimates of the approximate stellar mass are given for two isochrones; $10^6$\,yrs and $10^{7.5}$\,yrs. 
    }
    \centering
    \begin{tabular}{llrrll} \hline\hline
    Spectral& \teff &  M$_V$ & EW   & \multicolumn{2}{l}{Mass ($M_\odot$)}\\
    type    &  K    &   mag  & (m\AA) & 6.0 & 7.5 \\ \hline
    B2\,V   & 21500 & $-1.7$ & 30  & 8.8 & 6.9\\
    B2.5\,V & 20000 & $-1.1$ & 20  & 7.0 & 6.1\\
    B3\,V   & 16500 & $-0.6$ & 15  & 5.9 & 5.3\\
    B5\,V   & 15000 & +0.3   & 8.6 & 4.0 & 3.7\\
    B8\,V   & 13500 & +0.6   & 6.0 & 3.4 & 3.2\\
    B9\,V   & 12000 & +1.2   & 3.4 & 2.5 &2.5\\
    \hline
    \end{tabular}
    \label{tab:C.1}
\end{table}

\section{Spectral energy distribution}
\label{appendix:sed}

Sources of optical photometry listed in Table~\ref{tab:photometry} include the MCPS catalog of \citet{zaritsky}, \citet[][Evans]{evans2006} and $Gaia$ \citep{gaiabrown}.
This source lies within the boundaries of the MACHO survey of the LMC \citep{alcock} and was monitored for a period of almost 7 years.  
Mean magnitudes in the Kron-Cousins system of $V=15.50\pm0.04$ and $R=15.64\pm0.04$ are derived from the instrumental magnitudes using the transformations and zero-points described in \citet{alcock1999}.
Near-IR data are from the IRSF Magellanic Clouds catalog \citep{irsf}, which we prefer to 2MASS as the latter only contains upper limits for the $H$ and $K_s$ magnitudes, and the uncertainty of $J$ is 0.1$^m$. The mid-IR magnitudes are from the Spitzer SAGE survey \citep{sage}, and are taken from the catalog of \citet{bonanoslmc}. The source was not detected longwards of the 4.5$\mu$m band. 

\begin{figure}
    \centering
    \includegraphics[width=\hsize]{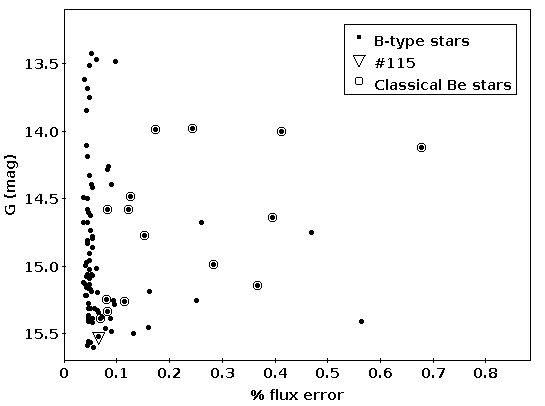}
    \caption{Gaia $G$ magnitude is plotted against its \% flux error, as an indication of variability for the B-type stars observed in NGC\,2004 by \citet{evans2006}. Star \#115 (inverted triangle, lower left) is among the least variable stars in this plot. Candidate classical Be stars are circled \citep[from][]{dunstall2011}.}
    \label{fig:gaia}
\end{figure}

\begin{figure}
    \centering
    \includegraphics[width=.8\hsize]{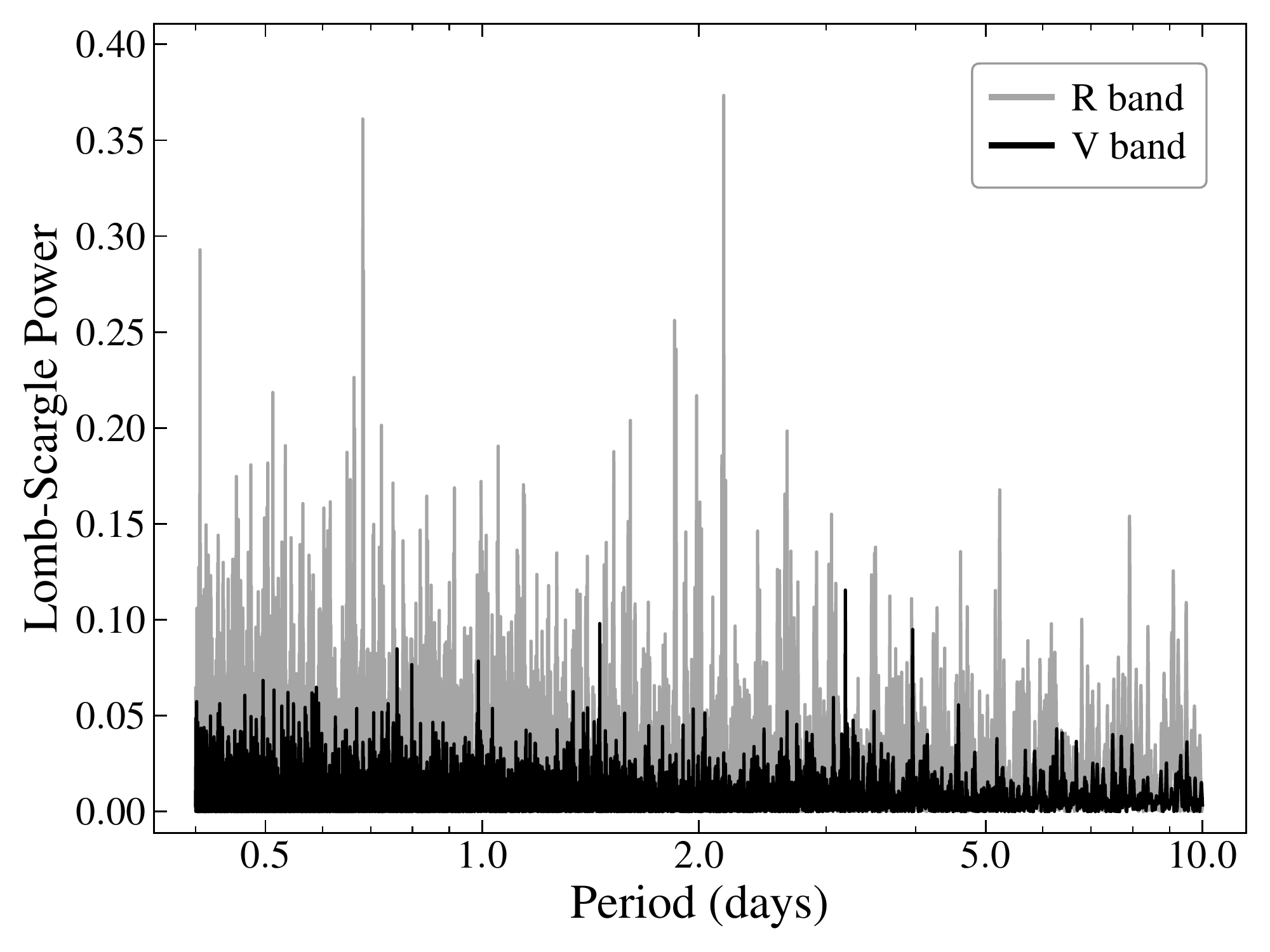}
    \caption{Zoom of Lomb-Scargle periodograms of the MACHO $V$ and $R$ band photometry below 10 days illustrating the lack of a significant peak in either band.}
    \label{fig:lomb}
\end{figure}

\citet{s11} used the MACHO data to investigate the field of NGC\,2004 in a search for variable stars, finding that the light curves of NGC\,2004\#115, in the MACHO $V$ and $R$ bands, are remarkably constant. 
A Lomb-Scargle test on both MACHO $V$ and $R$ bands does not find signs of the spectroscopic period. Two peaks above the 0.1\% false-alarm-probability level are found at 3.2 and 2.2 days in the case of the $V$ and $R$ bands respectively. However, the discrepancy between these two periods and the shape of the periodograms suggest that these are spurious peaks, Fig.~\ref{fig:lomb}. Folding the photometry data to the spectroscopic period (Fig.~\ref{fig:macho}) we estimate 1-$\sigma$ limiting peak-to-peak amplitudes of 0.002 mag, in the case of the $V$ band, and 0.006 mag for the $R$ band. These constrain the amplitude of possible ellipsoidal light variations to less than 0.005$^m$ in $V$ and 0.007$^m$ in $R$ in amplitude, or around 0.5--0.6\%.  The lack of variability of the system is also reflected in the $Gaia$ EDR3 data. While the project has yet to release all the epoch photometry we note that its $G$-band flux is given as 1.2e4\,$e^-$\,s$^{-1}$, with a 1-$\sigma$ error of 10.73\,$e^-$\,s$^{-1}$ based on 459 observations. In fact, according to Gaia data, NGC\,2004\#115 is one of the least variable B-type stars observed in NGC\,2004 by the FSMS, as illustrated in Figure \ref{fig:gaia}.
While H$\alpha$ displays significant profile variability, this has negligible impact on broad-band magnitudes.

The field of this system was also the target of two pointed XMM-Newton observations of $\sim$5000 seconds each, on 2007-01-28 and 2011-12-20, although there are no obvious X-ray sources in either data-set near its position. 
We used a 15\arcsec radius aperture to obtain 2-$\sigma$ upper limits of 0.0066 and 0.0034 counts/sec respectively in the 0.2--12 keV band. However, combining the images to get a total exposure time of 10.2 ks results in an upper limit of 0.0055 counts/s, as the higher background in the first observation dominates the combined image.
Converting the lowest of these limits into flux using using a power-law of slope of two, and hydrogen column density of N$_{\rm H}$=$2.7\times10^{21}$ cm$^{-2}$, yields limits for the absorbed flux of $1.13\times10^{-14}$ ergs\,cm$^{-2}$\,s$^{-1}$, and for the unabsorbed flux, $1.78\times10^{-14}$ ergs\,cm$^{-2}$\,s$^{-1}$.  This implies an X-ray flux from the system of less than $5.33\times10^{33}$ ergs\,s$^{-1}$ in the 0.2--12\,keV band.

Given the possible composite nature of the SED we adopted the mean extinction for the cluster of $E(B-V)$=0.08 \citep{caloi,cassatella} as appropriate for NGC\,2004\#115, and corrected the photometry for extinction using a combined average MW and LMC extinction curve \citep{gordon2003}. Using the filter function taken from \citet{bessell1990}, filter weighted model $V$-band fluxes are normalized to the observed $V$-band flux to yield the ratio of stellar radius to distance ($R/d$). From Table \ref{tab:photometry} it is apparent there are some small zero point offsets in the various $V$ magnitudes (more so in the $B$ band), hence we simply adopt the MACHO result in determining the radius in sections 3 and 4. The SED, illustrated in Fig.~\ref{fig:sed}, exhibits only a marginal IR excess at 4.5$\mu$m when compared with a model having no disk contribution. 

\begin{table}
    \caption{Photometric passbands with effective wavelengths and zero points used to convert the listed magnitudes to flux are from \citet{bonanoslmc,bessell98}. For the very broad $Gaia$ filters we adopt the pivot wavelengths of \citet{weiler2018}. The sources of the photometry are described in Appendix \ref{appendix:sed}.}
    \centering
    \begin{tabular}{llllll}\hline\hline
    Filter & $\lambda_{eff}$ ($\mu$m) & Zpt. (Jy) & Mag. & $\pm$ & Source \\\hline
$U$    & 0.366 &1790. &  14.73 &  0.07  &MCPS \\
$B$    & 0.438 &4063. &  15.36 &  0.16  &MCPS \\
$B$    & 0.438 &4063  &  15.25 &  0.07  &Evans\\
$Bp$   & 0.509 &3535. &  15.37 &  0.001 &Gaia \\
$V$    & 0.545 &3636. &  15.53 &  0.05  &MCPS \\
$V$    & 0.545 &3636. &  15.44 &  0.06  &Evans\\ 
$V$    & 0.545 &3636. &  15.50 &  0.04  &MACHO\\ 
$R$    & 0.641 &3064. &  15.64 &  0.04  &MACHO\\
$G$    & 0.623 &3296. &  15.51 &  0.001 &Gaia \\
$I$    & 0.798 &2416. &  15.75 &  0.03  &MCPS \\
$Rp$   & 0.777 &2620. &  15.64 &  0.001 &Gaia \\ 
$J$    & 1.220 &1589. &  15.78 &  0.01  &IRSF \\
$H$    & 1.630 &1021. &  15.79 &  0.02  &IRSF \\
$K_s$   & 2.120 &676. &  15.78 &  0.06  &IRSF \\
$[3.6]$& 3.55  &280.9 &  15.91 &  0.04  &SAGE \\
$[4.5]$& 4.49  &179.7 &  15.75 &  0.07  &SAGE \\ \hline
    \end{tabular}
    \label{tab:photometry}
\end{table}

\end{appendix}

\end{document}